\documentclass[aip,pof,natbib]{revtex4-1}

\usepackage{graphicx}
\usepackage{dcolumn}
\usepackage{bm}
\usepackage[utf8]{inputenc}
\usepackage[T1]{fontenc}
\usepackage{mathptmx}
\usepackage{etoolbox}
\usepackage{xcolor}
\usepackage{url}
\usepackage[normalem]{ulem}
\usepackage{subfigure}
\usepackage{amsmath}
\usepackage{booktabs}
\usepackage{epstopdf, epsfig}

\makeatletter
\def\@email#1#2{%
 \endgroup
 \patchcmd{\titleblock@produce}
  {\frontmatter@RRAPformat}
  {\frontmatter@RRAPformat{\produce@RRAP{*#1\href{mailto:#2}{#2}}}\frontmatter@RRAPformat}
  {}{}
}%
\makeatother
\begin{document}


\title[Separation-induced transition in a low pressure turbine under varying compressibility]{Separation-induced transition in a low pressure turbine under varying compressibility\\}

\author{Priya Pal}%

\author{Abhijeet Guha}%

\author{Aditi Sengupta}
\email{aditi@iitism.ac.in}

\affiliation{Department of Mechanical Engineering, Indian Institute of Technology Dhanbad, Jharkhand-826 004, India.
}%


\date{\today}

\begin{abstract}
The present study investigates influence of compressibility on separation-induced transition in a low-pressure turbine (LPT) cascade using high-fidelity numerical simulations of the T106A blade. Simulations are performed for inlet Mach numbers, $M_s$ ranging from 0.15 to 0.35 at a fixed Reynolds number and high incidence, representative of off-design LPT operation. A dispersion-relation-preserving numerical framework is employed to accurately capture instability waves, separation bubbles, and separation-induced transition to turbulence. A comprehensive analysis is carried out using surface pressure and skin-friction distributions, boundary-layer integral parameters, spectral analyses, and budgets of compressible enstrophy. Increasing $M_s$ systematically reduces streamwise extent of both leading-edge and trailing-edge separation bubbles and promotes earlier transition and reattachment, consistent with trends observed under increased free-stream disturbances. Despite shorter separation regions, suction-side momentum thickness at trailing edge increases by nearly 350\% from $M_s = 0.15$ to 0.35, indicating higher profile losses at elevated $M_s$. Spectral analyses demonstrate a redistribution of turbulent spatial and temporal scales, with energy injection occurring at progressively larger scales as $M_s$ increases. Flow-field visualizations reveal a transition pathway that shifts from two-dimensional spanwise rolls and intermittent turbulent spots at low $M_s$ to streak-dominated, bypass-like transition at higher $M_s$. Compressible enstrophy budgets reveal that viscous–compressible coupling and baroclinic mechanisms dominate vorticity dynamics. Thus, compressibility fundamentally alters transition mechanisms and loss generation in LPT flows, underscoring the need for vorticity- and enstrophy-based analyses in addition to conventional boundary-layer characterization.

\end{abstract}

\maketitle

\section{Introduction \label{sec1}}

Low-pressure turbines (LPTs) play a critical role in determining the overall efficiency of modern aero-engines and power-generation systems \cite{halstead1997boundary}. In pursuit of reduced fuel consumption and emissions, LPT blades are increasingly designed to operate at high lift, low Reynolds numbers ($Re$), and under off-design incidence conditions. Under such operating regimes, the suction-surface boundary layer is prone to laminar separation, leading to separation-induced transition and the formation of laminar separation bubbles \cite{hammer2018large}. The dynamics of these bubbles strongly influence loss generation, wake characteristics, and stage efficiency, making separation-induced transition a central problem in LPT aerodynamics. Banieghbal {\it et al.} \cite{banieghbal1996wake} showed that suction surface losses have a 60\% contibution from boundary layer losses, which are significantly influenced by the external blade environment. To enhance engine efficiency, designers seek to explain underlying flow physics governing the transition process in a flow characterized by an adverse pressure gradient, as encountered in LPTs.

Early experimental and theoretical studies established that, unlike classical Tollmien–Schlichting-wave-driven transition, LPT transition is often governed by the instability and breakdown of separated shear layers \cite{coull2012predicting}. A seminal work \cite{mayle1991role} highlighted the dominant role of separation-induced transition in gas turbines, while subsequent cascade experiments \cite{stadtmuller2001investigation} and numerical studies \cite{wissink2006influence, michelassi2015compressible} on representative LPT blades such as the T106A geometry provided detailed insight into separation bubble topology, transition onset, and reattachment behavior under varying $Re$ and free-stream turbulence levels \cite{volino2001measurements, simoni2015off, sengupta2020effects}. The T106A blade is a representative \lq high-lift' LPT airfoil, characterized by a comparatively smooth loading distribution along the chord, in contrast to the abrupt pressure gradients typically associated with flat-plate idealizations. It corresponds to a mid-span section of the Pratt \& Whitney PW2037 LPT and has therefore emerged as a canonical configuration for studying transitional LPT flows. Owing to its practical relevance, the T106A geometry has been extensively examined through both experimental measurements \cite{stadtmuller2001test} and numerical simulations \cite{wissink2003dns}. Early numerical investigations \cite{ranjan2014direct} of flow through a T106A blade passage at $Re \approx 51800$ revealed complex suction-surface dynamics, including the formation of multiple laminar separation bubbles and repeated laminar–turbulent transition cycles. Subsequent DNS studies \cite{de2023effects} focused on unsteady rotor–stator interactions, demonstrating how periodically impinging wakes modify the instantaneous flow field and transition behavior within the cascade. A broader parametric investigation by Michelassi et al. \cite{michelassi2015compressible}, examined the combined influence of background turbulence intensity, wake-passing frequency, and $Re$. The analysis highlighted limitations of the Boussinesq eddy-viscosity hypothesis, with the largest discrepancies observed near the trailing edge. The sensitivity of T106A flows to geometric and operational variations has also been explored. Pichler et al. \cite{pichler2017highly} investigated the effect of stator–rotor axial spacing, considering gaps ranging from 21.5\% to 43\% of the rotor chord, and reported higher overall losses for smaller gaps due to enhanced amplification of wake-induced turbulence. The impact of surface roughness was examined using large-eddy simulations \cite{hammer2018large}, employing both forcing-based roughness models and immersed boundary techniques. Their results showed that realistic as-cast roughness promotes streak formation and leads to pronounced spanwise non-uniformity in transition onset. Additional studies have addressed the role of free-stream turbulence, demonstrating that at a chord $Re$ of 30,000, elevated turbulence levels can induce Klebanoff streaks and suppress suction-side separation \cite{gross2022numerical}. More recently, spectral proper orthogonal decomposition has been applied to the T106A cascade \cite{fiore2023t106}, revealing that upstream wakes selectively energize dominant boundary-layer modes and intensify turbulent structures in the blade wake downstream of the trailing edge.

While the influence of $Re$, free-stream disturbances, roughness \cite{sengupta2017roughness, hammer2018large}, aeroelasticity \cite{nakhchi2021direct, sengupta2020effectsa} and unsteady wakes \cite{wissink2006influence, michelassi2002analysis} on LPT transition has been extensively studied, the role of compressibility remains comparatively underexplored, particularly in the context of separation-induced transition. Practical LPTs operate with Mach numbers typically ranging from 0.1 to 0.4, where compressibility effects, though moderate, can no longer be neglected. Existing studies have shown that increasing Mach number can suppress large separation bubbles and promote earlier transition \cite{fiore2021reynolds, sengupta2023compressibility, duan2023effects}, often drawing analogies with the effects of elevated free-stream turbulence. However, most of these studies have focused on time-averaged metrics on the suction surface boundary layer such as pressure and skin-friction distributions, offering limited insight into the underlying instability mechanisms.

Recent advances in high-fidelity numerical methods have made it possible to resolve the unsteady, three-dimensional dynamics of separation-induced LPT transition in compressible flows \cite{michelassi2015compressible, ranjan2014direct, sengupta2023compressibility}. Yet, the majority of prior work relied on turbulent kinetic energy–based interpretations, which are not always sufficient for compressible flows where vorticity generation can arise from baroclinic and dilatational mechanisms \cite{pirozzoli2011numerical}. Consequently, the vorticity and enstrophy dynamics governing compressible separation-induced transition in LPTs remain poorly understood. The present study addresses this gap by performing high-fidelity DNS of the T106A LPT cascade over a range of inlet Mach numbers while maintaining fixed $Re$ and incidence. In addition to conventional analyses based on surface pressure, skin friction, and boundary-layer integral parameters, the study employs space–time diagnostics, spectral analyses, turbulent kinetic energy distributions, and a compressible enstrophy transport equation \cite{suman2022novel} to quantify the mechanisms responsible for vorticity production and loss generation. By linking compressibility-induced changes in transition pathways to enstrophy budgets, this work provides a unified, physics-based interpretation of how moderate Mach number variations alter separation, transition, and aerodynamic losses in LPT flows. Through this analysis, the paper aims to (i) clarify the role of compressibility in separation-induced transition mechanisms, (ii) reconcile the apparent contradiction between reduced separation extent and increased profile losses at higher Mach numbers, and (iii) establish enstrophy-based diagnostics as a valuable tool for modeling and design of future low-pressure turbine blades.

The paper is formatted as follows: In section \ref{sec2}, the problem formulation for flow inside a T106A cascade passage is provided. This includes the governing equations, boundary conditions and numerical methodology adopted. Validation efforts with benchmark DNS data \cite{stadtmuller2001test, wissink2003dns} are also detailed. Section \ref{sec3} describes the results and discussion by considering instantaneous and time-averaged flow features for vorticity, its spectra and integrated boundary layer parameters. The budgets of turbulent kinetic energy and CETE are also explored for varying inlet Mach numbers, highlighting the role of compressibility on separation-induced transition mechanisms. The paper closes with a summary and conclusions in section \ref{sec4}.

\section{Problem formulation of the T106A cascade \label{sec2}}

\begin{figure*}
\centering
\includegraphics[width=.9\textwidth]{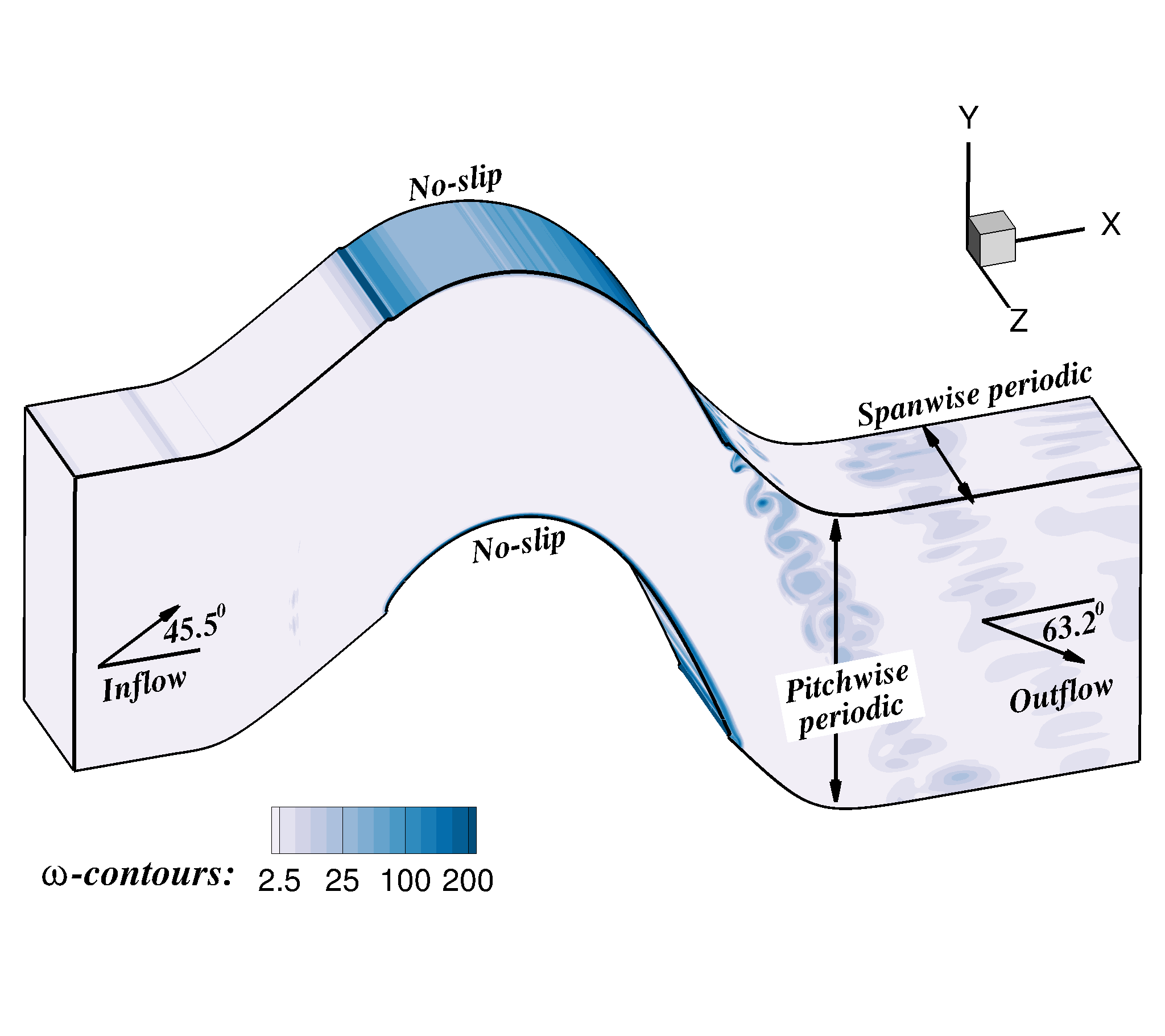}
\caption{Schematic of flow inside a T106A low pressure turbine blade passage coloured with magnitude of vorticity contours.}
\label{fig1}
\end{figure*}

The computational domain and the corresponding boundary conditions are illustrated in Fig.~\ref{fig1}. The present setup is motivated by experimental studies \cite{opoka2007transition, stadtmuller2001test} carried out on a linear T106A turbine blade cascade in a high-speed cascade wind tunnel at the Universität der Bundeswehr, Munich, Germany. These experiments were intended to provide reliable low-Reynolds-number benchmark data for the validation of numerical simulations. Owing to the comparatively lower aerodynamic loading of the T106A blade relative to ultra-high-lift blade configurations \cite{wissink2004dns}, the suction-side boundary layer remains relatively thin. A no-slip boundary condition is prescribed on the blade surface, while periodic boundary conditions are imposed in the pitchwise direction. The inlet boundary is positioned at a distance of 0.4 axial chord lengths upstream of the leading edge, and the outlet extends up to 0.5 axial chord lengths downstream of the trailing edge. A H-type structured grid consisting of $1251 \times 501 \times 256$ grid points in the streamwise, wall-normal, and spanwise directions, respectively, is employed. Of the 1251 streamwise points, 276 are distributed upstream of the leading edge, 751 along the blade surface, and 226 downstream of the trailing edge. The simulations are performed at a Reynolds number of $1.6 \times 10^{5}$, defined based on the true chord length $c$ and the exit velocity $U_{TE}$, consistent with the DNS study of Wissink \cite{wissink2003dns}. The blade exhibits a mid-loaded pressure distribution, with the peak suction located at $42\%\,S_0$, where $S_0$ denotes the suction-side surface length. The blade pitch is fixed at 0.9306. At the inflow boundary, the flow enters with an incidence angle of $\alpha_{in} = 45.5^\circ$, and an inlet Mach number $M_s$, corresponding to a Reynolds number of 60142.72. The outflow boundary is characterized by an exit flow angle of $\alpha_{out} = -63.2^\circ$ and an outlet Mach number $M_{out}$. To prevent spurious wave reflections at the inflow and outflow boundaries from contaminating the solution, characteristic-based boundary conditions formulated using Riemann invariants \cite{hirsch1990numerical} are applied.

The numerical simulations are performed by solving the unsteady 3D compressible NSE \cite{hoffmann2000computational} given as:

\begin{equation}
\frac{\partial \hat{Q}}{\partial t^*}+\frac{\partial \hat{E_c}}{\partial x^*}+\frac{\partial \hat{F_c}}{\partial y^*}+\frac{\partial \hat{G_c}}{\partial z^*}=\frac{\partial \hat{E_v}}{\partial x^*}+\frac{\partial \hat{F_v}}{\partial y^*}+\frac{\partial \hat{G_v}}{\partial z^*},
\label{ge1}
\end{equation}

\noindent for the conserved variables given as
\begin{equation}
\hat{Q} = [ \rho^*; \; \rho^* u^*; \; \rho^* v^*; \; \rho^* w^*; \; \rho^* e_t^* ]^T.
\label{ge2}
\end{equation}
	
\noindent The convective flux variables $\hat{E}_c$, $\hat{F}_c$ and $\hat{G}_c$ are given as

\begin{equation}
\hat{E}_c = [ \rho^* u^*; \; \rho^* u^{*2} + p^*; \; \rho^* u^* v^* ; \; \rho^* u^* w^*; \; (\rho^* e_t^* + p^*) u^*]^T,
\label{ge3}
\end{equation}
 
\begin{equation}
\hat{F}_c = [ \rho^* v^*; \; \rho^* u^* v^*; \; \rho^* v^{*2} + p^*; \; \rho^* v^* w^*; \; 
(\rho^* e_t^* + p^*) v^* ]^T,
\label{ge4}
\end{equation}

\begin{equation}
\hat{G}_c = [ \rho^* w^*; \; \rho^* u^* w^*; \; \rho^* v^* w^*; \; \rho^* w^{*2} + p^*; \; (\rho^* e_t^* + p^*) w^* ]^T,
\label{ge5}
\end{equation}
	
\noindent and the viscous flux vectors $\hat{E}_v$, $\hat{F}_v$ and $\hat{G}_v$ are given as:

\begin{equation}
\hat{E}_v = [ 0; \; \tau^*_{xx}; \; \tau^*_{xy}; \; \tau^*_{xz}; \; u^* \tau^*_{xx} + v^* \tau^*_{xy} + w^* \tau^*_{xz} - q^*_x ]^T,
\label{ge6}
\end{equation}

\begin{equation}
\hat{F}_v = [ 0; \; \tau^*_{yx}; \; \tau^*_{yy}; \; \tau^*_{yz}; \; u^* \tau^*_{yx} + v^* \tau^*_{yy} + w^* \tau^*_{yz} - q^*_y ]^T,
\label{ge7}
\end{equation}

\begin{equation}
\hat{G}_v = [ 0; \; \tau^*_{zx}; \; \tau^*_{zy}; \; \tau^*_{zz}; \; u^* \tau^*_{zx} + v^* \tau^*_{zy} + w^* \tau^*_{zz} - q^*_z ]^T.
\label{ge8}
\end{equation}
	
Flow variables $\rho^*$, $p^*$, $u^*$, $v^*$, $w^*$, $T^*$ and $e_t^*$ represent dimensional density, pressure, Cartesian components of velocity, the absolute temperature and specific internal energy, respectively. The stress tensor $\tau^*_{ij}$, for $i,j = 1$ to 3, is related to the rate of strains as,

\begin{equation}
\tau^*_{ij} = \tau^*_{ji} =  \mu^* \biggl( \frac{\partial v_{j}^*}{\partial x_{i}^*} +  \frac{\partial v_{i}^*}{\partial x_{j}^*} \biggr );
		\; \tau^*_{ij} = \biggl[ \mu^* \biggl (\frac{\partial v_{j}^*}{\partial x_{i}^*} +  \frac{\partial v_{i}^*}{\partial x_{j}^*} \biggr ) + \lambda^* \frac{\partial v_{i}^*}{\partial x_{i}^*}   \biggr ] \; {\rm for} \; i = j;
\label{stress}
\end{equation}
	
\noindent The specific heat capacity ($C_v$) and thermal conductivity ($\kappa$) are constants in this formulation. The heat conduction terms $q^*_i$ are given as: $q^*_{i} = - \kappa \frac{\partial T^*}{\partial x^*_{i}}$. Sutherland's law is used for evaluating the dynamic viscosity ($\mu^*$) as a function of temperature ($T^*$), and Stokes' hypothesis is implemented for the bulk viscosity calculation. Additionally, the ideal gas relation, $p^*=\rho^* R^*T^*$ is the gas equation of state, which defines specific energy, $e_t^*$. With summation convention of Einstein, it is formulated as
	
\begin{equation}
e_t^* =\frac{p^*}{\rho^*(\gamma-1)} + \frac{v_{i}^*v_{i}^*}{2}.
\label{ge13}
\end{equation}
	
\noindent The independent and dependent variables are nondimensionalized \cite{hoffmann2000computational} as follows,
	
\begin{equation*}
x_i = \frac{x^*_{i}}{L}, \; v_i = \frac{v^*_{i}}{U_s}, \;t = \frac{t^*U_s}{L}, \; \rho = \frac{\rho^*}{\rho_s}, \; p = \frac{p^*}{p_s},\; e_t = \frac{e^*_t}{U^2_s}, \; T = \frac{T^*}{T_s}, \; \mu = \frac{\mu^*}{\mu_s}
\end{equation*}
 
\noindent where $L$, $U_s$, $T_s$ and $\rho_s$ are the length, velocity, temperature, and density scales and $p_s = \rho_sR^*T_s$, is the characteristic pressure obtained with the reference temperature and density. Non-dimensionalization is performed using temperature and density scales, specifically $T_s = 300K$ and $\rho_s = 1.177 kg/m^3$ and $L$ is taken as the chord of the T106A blade. Five numerical simulations are conducted by varying the Mach number, $M_s$ at the inflow to the T106A blade passage. The particulars of each test case are outlined in Table \ref{tab1}. Each simulation utilizes 96 cores over 720 compute hours, resulting in a total of 69120 core hours per test case. Initial transients in the flow are eliminated by simulating five to six through-flows. Additionally, for time-averaging the flow field, an extra 10-12 through-flows are computed, and time-resolved data is stored.

\begin{table}[!ht]
\centering
\caption{Details of the test cases simulated.}
\vspace{1mm}
\begin{tabular}{|c| c|}
\hline
 Case & Description  \\ [0.5ex]
 \hline\hline
    Case-1 & Test case with inlet Mach number, $M_s$ = 0.15 and outlet Mach number, $M_{out} = 0.404$.  \\
    Case-2 & Test case with inlet Mach number, $M_s$ = 0.20 and outlet Mach number, $M_{out} = 0.534$.   \\
    Case-3 & Test case with inlet Mach number, $M_s$ = 0.25 and outlet Mach number, $M_{out} = 0.668$.   \\
    Case-4 & Test case with inlet Mach number, $M_s$ = 0.30 and outlet Mach number, $M_{out} = 0.801$.   \\
    Case-5 & Test case with inlet Mach number, $M_s$ = 0.35 and outlet Mach number, $M_{out} = 0.935$. \\           [1ex]
 \hline
\end{tabular}

\label{tab1}
\end{table}

\subsection{Numerical methodology and validation}

\noindent The present high-fidelity simulations employ a dispersion-relation-preserving (DRP) compact finite-difference framework \cite{sagaut2023global}, specifically optimized to maintain neutral stability for the range of wavenumbers and frequencies encountered in transitional LPT flows. Spatial discretization of the convective fluxes is performed using the OUCS3 scheme, which ensures that the numerical dispersion relation closely matches the physical one. This capability is crucial for accurately resolving the wave-dominated unsteady phenomena that develop over the suction surface of the T106A blade, where a cascade of separation bubbles and non-modal disturbances govern the transition process \cite{sengupta2020effects}. As documented in earlier studies \cite{sengupta2020nonmodal, sundaram2020effects}, the dynamics of these separation bubbles are extremely sensitive to numerical dissipation and dispersion. The DRP characteristics of OUCS3 therefore play a central role in capturing the correct phase speed, group velocity, numerical amplification, and interaction of the instability waves that trigger transition under compressible conditions. To further enhance numerical stability while preserving accuracy, a one-dimensional explicit filter is applied independently in the $x$, $y$, and $z$ directions using a filter coefficient of 0.47. This filtering strategy removes high-frequency spurious modes without compromising the physically relevant scales associated with the separation-bubble dynamics. The viscous fluxes are discretized using a second-order central-difference ($CD_2$) scheme. Unlike convective terms that transport waves and therefore demand careful dispersion control, viscous terms inherently contain physical dissipation. Thus, a simple central scheme —which is non-dissipative numerically — is sufficient and appropriate. Adding numerical dissipation would distort the physical diffusion already present in the equations. Temporal advancement is performed using the OCRK3 scheme, an optimized variant of the three-stage Runge–Kutta method specifically designed to offer improved DRP properties compared to the classical four-stage, fourth-order Runge–Kutta algorithm \cite{sagaut2023global}. The use of OCRK3 ensures that numerical dispersion errors remain minimal during time marching, which is essential for faithful representation of unsteady wave propagation on the suction surface. Finally, the time step is chosen to maintain neutral stability of the combined spatial and temporal discretization. This ensures that physically important flow structures — such as convective instabilities, streak-induced instabilities, Kelvin-Helmholtz roll-ups, and the evolving 3D separation bubbles — are neither artificially damped nor spuriously amplified by the numerics.

\begin{figure*}
\centering
\includegraphics[width=.8\textwidth ]{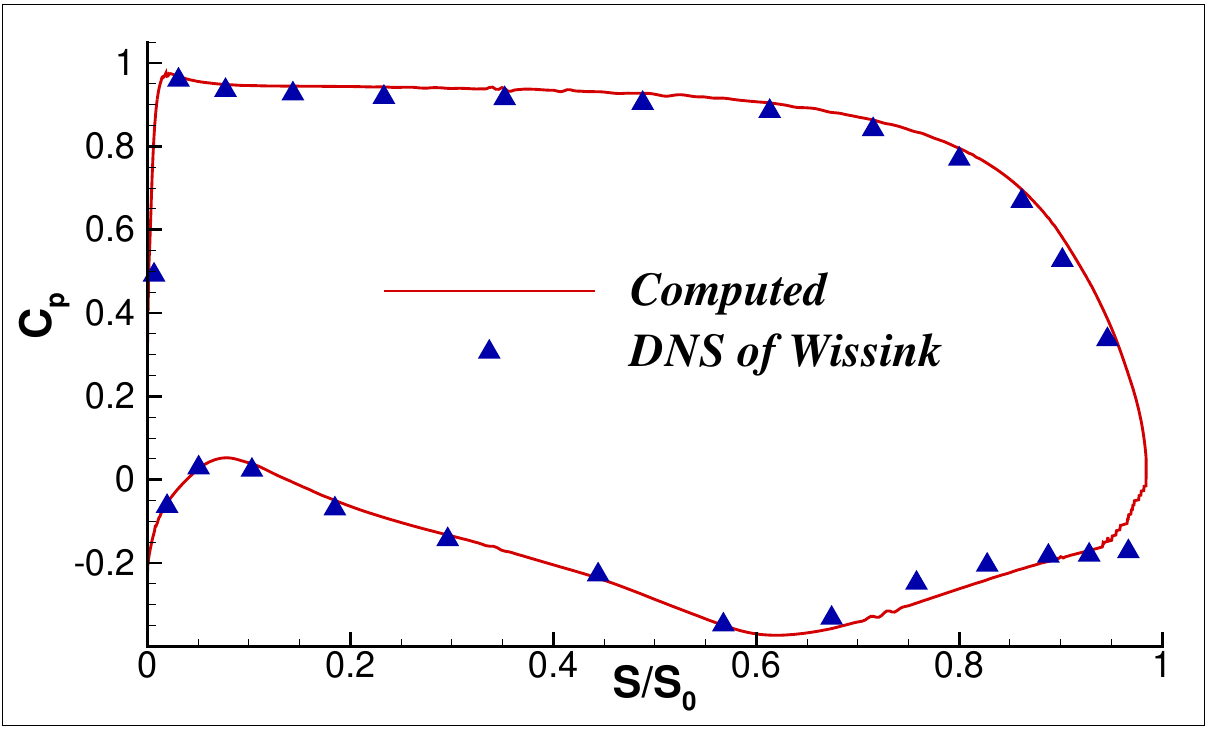}
\caption{Comparison of computed streamwise variation of time-averaged $C_p$ with the DNS results of Wissink \cite{wissink2003dns} for $M_s = 0.15$. The top line represents the pressure distribution on the pressure surface, while the bottom line is the suction surface's $C_p$.}
\label{fig2}
\end{figure*}

Figure \ref{fig2} presents the time-averaged distribution of the wall static pressure coefficient, $C_p$ plotted against the normalized surface coordinate, $S/S_0$ for the T106A blade at an inlet Mach number, $M_s = 0.15$. The curve is shown separately for the pressure and suction surfaces, together with benchmark data from Stadtmüller’s \cite{stadtmuller2001test} experiments and Wissink’s DNS \cite{wissink2003dns} for nominally disturbance-free inflow. Overall, the computed $C_p$ variation closely matches the DNS predictions on both blade surfaces, indicating that the present numerical approach captures the essential pressure loading of the T106A profile. On the pressure surface, the flow experiences a strong favorable pressure gradient as it accelerates toward the trailing edge. This is reflected in a monotonic decrease in $C_p$ along the streamwise coordinate. The close match between the present results, DNS, and experiments reflects the largely attached and quasi-two-dimensional nature of the pressure-side flow, which is comparatively insensitive to compressibility and inflow-angle variations. This agreement serves as a consistency check for the numerical framework.

The suction surface shows a more pronounced acceleration of the flow, visible as a sharper drop in $C_p$ as the flow approaches the suction-side peak. This region is sensitive to both Mach number and inflow angle, and therefore the comparisons show some key features: the present simulation reproduces the suction-side $C_p$ behavior from Wissink’s DNS \cite{wissink2003dns} very well, particularly up to and around the point of maximum acceleration. This indicates that the blade loading and suction-side pressure gradient are accurately captured under the assumed inflow conditions. Stadtmüller’s experiment \cite{stadtmuller2001test} imposed a maximum isentropic Mach number of 0.48, whereas the DNS and the present simulation operate at different effective Mach levels. Because the suction-side acceleration in the experiment is capped by this Mach limit, the experimental $C_p$ peak differs from both DNS and the present simulation. This explains the slight mismatch in regions of strong suction-side acceleration, where compressibility effects are most influential. The original experiments used an inflow angle of $37.7^{\circ}$, whereas both the DNS and the present simulations adopt $\alpha_{in} = 45.5^{\circ}$. Later RANS and 3D hot-wire measurements by Stadtmüller \cite{stadtmuller2001investigation} showed that the actual inflow angle was closer to $45.5^{\circ}$, likely due to the wake generator placed upstream of the cascade, which shifts the effective flow direction.
This correction reconciles the DNS and the current simulations with the true physical inflow experienced by the blade.

Distinct features in the computed and DNS $C_p$ distributions arise from laminar separation bubbles on the suction surface of the blade surface. The small plateau in $C_p$ near $S/S_0 \approx 0.05$ corresponds to a short separation bubble at the leading edge. Similar behavior has been documented in prior DNS studies \cite{wissink2003dns, wissink2006influence, michelassi2015compressible}. The deviations between the computed $C_p$ and the benchmark data in this region can be attributed to differences in the onset, size, and reattachment location of this bubble. A more persistent plateau in $C_p$ appears closer to the trailing edge, and this trend is reproduced both in the present results and the DNS. This plateau indicates the presence of a longer separation bubble extending toward the blade’s aft portion \cite{michelassi2015compressible, wissink2003dns}. The trailing-edge bubble influences the lift distribution and modifies the wake thickness, making its accurate resolution critical for LPT performance predictions.

\section{Results and discussion \label{sec3}}

This section presents a detailed analysis of the three-dimensional flow physics governing compressible, separation-induced transition over the T106A LPT blade. To elucidate the spatio-temporal evolution of unsteady structures responsible for transition, iso-surfaces of the Q-criterion and spanwise vorticity are examined, revealing the formation, breakdown, and convection of vortical packets originating from the suction-side separation bubble. The underlying dynamics are further quantified using space–time plots of wall-normal vorticity, enabling identification of characteristic temporal and streamwise propagation scales. Key flow-field statistics — including boundary-layer growth on the suction surface, along with the turbulent kinetic energy (TKE) and enstrophy budgets — are subsequently evaluated to establish the relative contributions of production, diffusion, vortex stretching, dissipation, and compressibility effects throughout the transitional process. Together, these diagnostics provide a comprehensive understanding of the mechanisms driving three-dimensional transition under the present compressible conditions.

\subsection{Vorticity dynamics at different inlet Mach numbers}

\begin{figure*}
\centering
\includegraphics[width=.9\textwidth ]{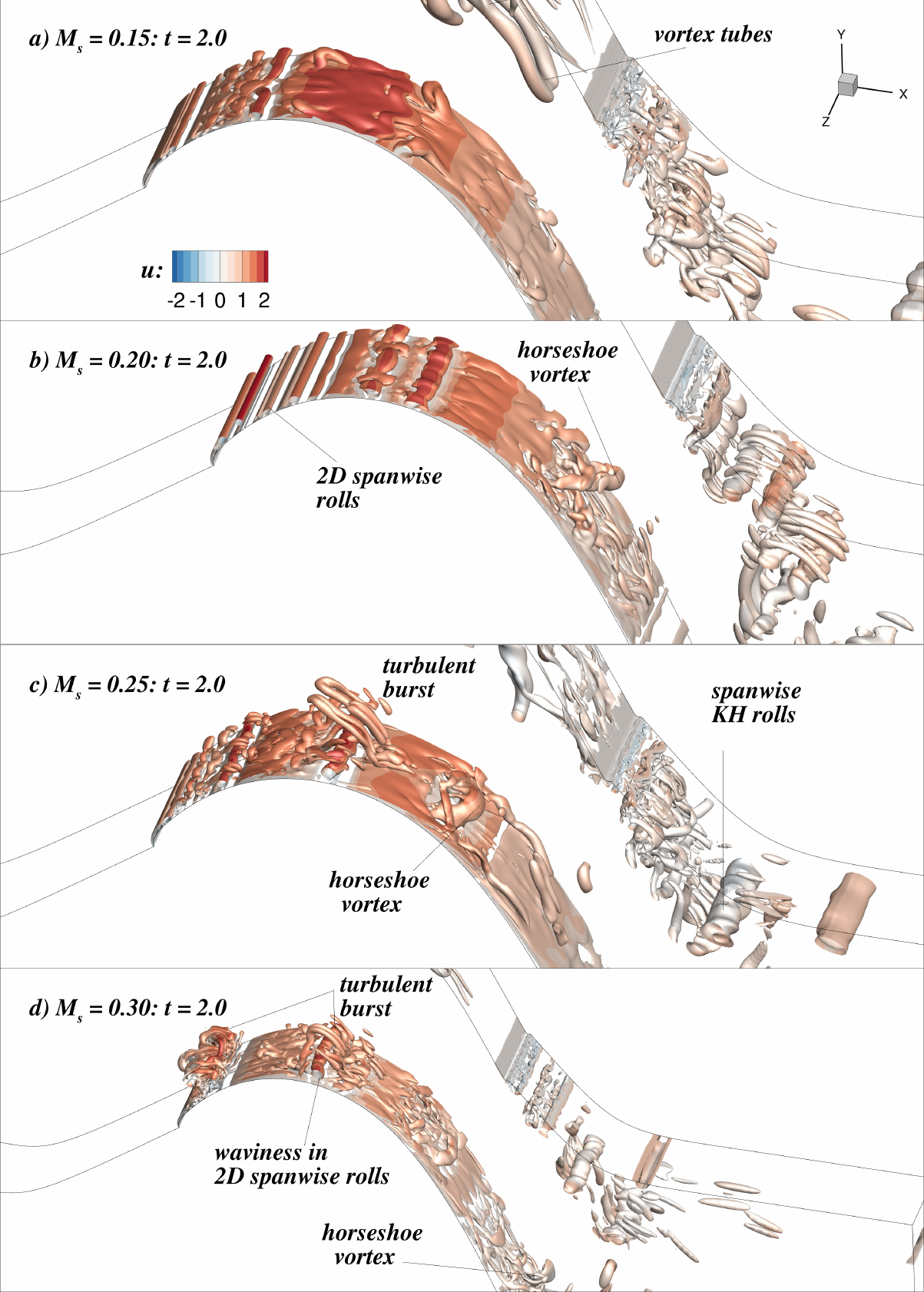}
\caption{Iso-surfaces of Q-criterion, $Q = 50$ colored with the streamwise velocity, $u$ for test cases with $M_s = 0.15$, 0.20, 0.25 and 0.30, showing the pertinent coherent structures.}
\label{fig3}
\end{figure*}

Figure \ref{fig3} illustrates the iso-contours of the Q-criterion for four inlet Mach numbers, $M_s = 0.15$, 0.20, 0.25 and 0.30, colored with the streamwise velocity, $u$. These visualizations highlight the evolution of coherent vortical structures on both the pressure and suction surfaces of the T106A blade as compressibility effects intensify.

At the lowest Mach number, $M_s = 0.15$ in Fig. \ref{fig3}(a), the flow on the pressure surface is characterized by the formation of elongated vortex tubes that remain attached over a significant portion of the surface. These tubes eventually undergo periodic distortion and detach from the trailing edge, leading to vortex shedding into the wake. This behavior reflects the relatively mild adverse pressure gradients on the pressure side and the dominance of convective instabilities at low compressibility. On the suction surface, the Q-criterion reveals the emergence of nearly two-dimensional spanwise vortical rolls originating within the separation bubble. These rolls grow, tilt, and deform downstream, introducing spanwise waviness to the separated shear layer. Their nonlinear interaction leads to the breakdown of the shear layer into localized turbulent spots, marking the onset of three-dimensional transition. This mechanism is consistent with the Kelvin–Helmholtz–type roll-up followed by secondary instabilities typically observed in LPT separation-induced transition \cite{sengupta2024separation, licari2011modeling}. As the inlet $M_s$ increases to 0.20 in Fig. \ref{fig3}(b), the suction-side dynamics become more complex. In addition to the Kelvin–Helmholtz-type spanwise rolls, the shear layer begins to organize into rows of hairpin vortices, which bridge the gap between the initial 2D roll-up and fully turbulent breakdown. These hairpin packets amplify the spanwise modulation of the separated shear layer, accelerating the formation of localized turbulent spots \cite{adrian2007hairpin, sundaram2020effects}. At $M_s = 0.25$ in Fig. \ref{fig3}(c), compressibility effects trigger an even earlier onset of three-dimensionality. The suction surface exhibits turbulent bursts and horseshoe vortices downstream of the leading edge. These structures indicate a more abrupt transition pathway, bypassing the orderly formation of 2D vortex rolls and hairpin trains \cite{matsubara2001disturbance} observed at lower Mach numbers. The flow becomes highly three-dimensional, with vortical packets growing and breaking down over a shorter streamwise distance. For the highest Mach number, $M_s = 0.30$ in Fig. \ref{fig3}(d), the suction surface shows the presence of multiple turbulent burst events originating almost immediately downstream of the leading edge. These bursts are of lower amplitude compared to the intense events for $M_s = 0.25$, suggesting a rapid but more distributed transition process. Notably, the classical 2D spanwise rolls seen at lower Mach numbers are absent. Instead, compressibility effects promote instantaneous spanwise waviness of the shear layer right from the leading edge, causing the flow to progress directly into a transitional stage without the intermediate stages of 2D roll-up or hairpin organization \cite{mohan2021influence}.

\begin{figure*}
\centering
\includegraphics[width=.9\textwidth]{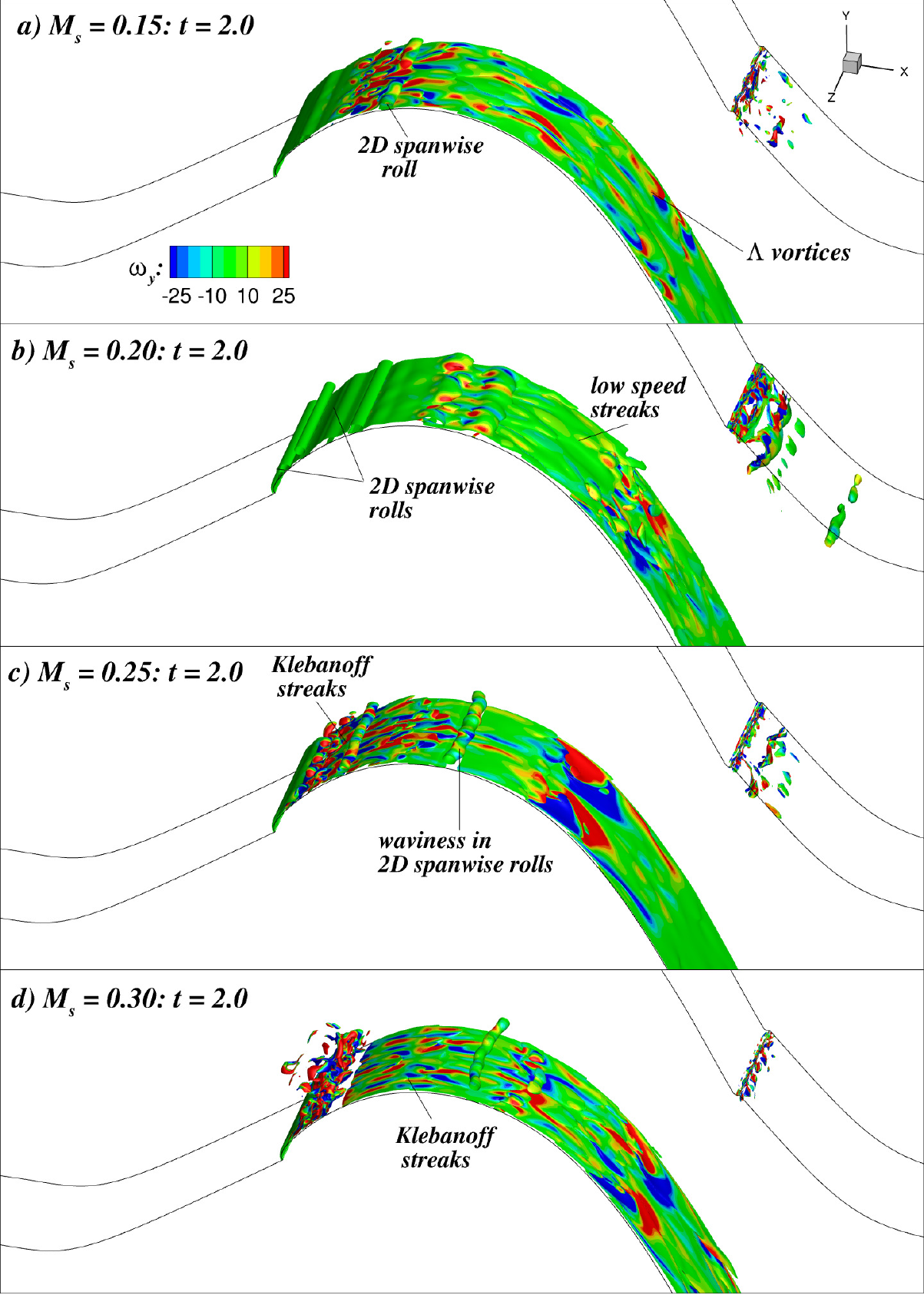}
\caption{Iso-surfaces of spanwise vorticity, $\omega_z = -50$ colored with wall-normal vorticity, $\omega_y$ for test cases with $M_s = 0.15$, 0.20, 0.25 and 0.30, showing the relevant vortical structures.}
\label{fig4}
\end{figure*}

Figure  \ref{fig4} presents iso-contours of spanwise vorticity ($\omega_z$) shaded using wall-normal vorticity ($\omega_y$) for inlet Mach numbers $M_s = 0.15$, 0.20, 0.25, and 0.30. The visualization highlights the evolution of vortical structures on the suction surface and the mechanisms leading to transition for varying compressibility levels. At the lowest Mach number in Fig. \ref{fig4}(a), the flow retains a predominantly two-dimensional character over a significant portion of the chord. Distinct 2D spanwise rolls are evident, with $\Lambda$-vortices forming closer to the trailing edge as the amplitude of disturbances grows downstream. The wall-normal vorticity contours reveal alternating-sign counter-rotating vortices in the aft portion of the blade, consistent with classical TS wave amplification and late-stage $\Lambda$-vortex formation \cite{adrian2007hairpin}. Increasing Mach number to 0.20 in Fig. \ref{fig4}(b) introduces enhanced disturbance growth and earlier three-dimensionality. The initially coherent 2D spanwise rolls gradually acquire spanwise waviness, signaling the onset of secondary instability. These undulations generate $\Lambda$-vortices that subsequently break down into low-speed streaks, indicating the emergence of near-wall streamwise elongation \cite{doering1995applied} and incipient turbulent activity. At moderate Mach number of 0.25 in Fig. \ref{fig4}(c), the effect of compressibility-driven destabilization is more prominent with 2D rolls breaking down rapidly, within a short distance from the leading edge, into alternately signed separation bubble-induced streaks \cite{klebanoff1962three}. These streaks grow in amplitude and eventually collapse into two dominant counter-rotating vortices near the trailing edge, marking a more abrupt and energetic transition scenario. At the highest Mach number of $M_s = 0.30$ in Fig. \ref{fig4}(d), the flow shows strong streaks immediately downstream of the leading edge, reflecting a highly receptive boundary layer and amplified disturbance environment. These streaks persist over much of the suction surface and exhibit intermittent turbulent bursts, demonstrating a breakdown pathway dominated by bypass-type mechanisms rather than classical TS dynamics \cite{sengupta2020nonmodal}. The early loss of spanwise coherence and frequent bursts highlight the intensified instability induced by higher Mach number inflow \cite{sandham2016effects}.

\begin{figure*}
\centering
\includegraphics[width=.9\textwidth]{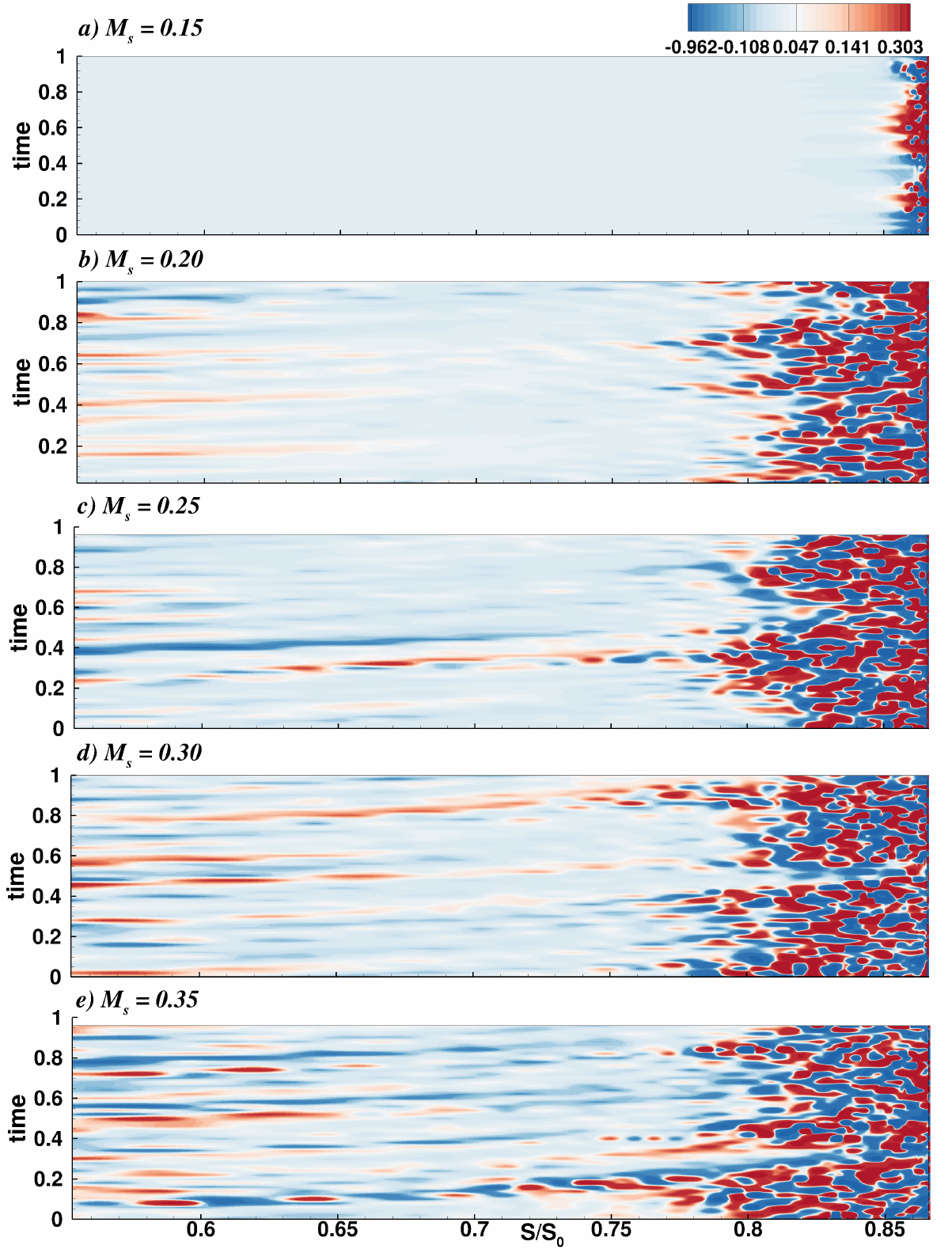}
\caption{Space-time plot of $\omega_y$ with indicated contour levels for test cases with (a) $M_s = 0.15$, (b) $M_s = 0.20$, (c) $M_s = 0.25$, (d) $M_s = 0.30$ and (e) $M_s = 0.35$ .}
\label{fig5}
\end{figure*}

The dynamic behaviour of the separated shear layer and the related dominant frequencies can be detected using space-time plots. It is one of the most insightful diagnostics for separation-induced transition, which helps to determine when, where, and how disturbances amplify and lead to breakdown. During the simulations, 3D instantaneous flow field data has been saved for every 50 time-steps, resulting in 200 snapshots. Using this data, space-time plots of $\omega_y$ have been shown for $M_s$ ranging from 0.15 to 0.35 in Fig. \ref{fig5}. At the lowest Mach number of $M_s = 0.15$ in Fig. \ref{fig5}(a), the space–time diagram shows a quiescent flow field extending over most of the suction surface. No significant wave activity or periodic patterns appear upstream, indicating a weakly unstable separated shear layer. Only near the aft portion of the blade ($S/S_0 > 0.8$) do we observe intermittent, high-intensity patches of wall-normal vorticity, signalling the sporadic emergence of turbulent spots. This behaviour is consistent with a delayed, weak KH-type instability and a long laminar separation bubble that remains mostly tranquil until close to the trailing edge. For $M_s = 0.20$ in Fig. \ref{fig5}(b), distinct streamwise streaks appear in the space–time plot for $S/S_0 > 0.5$, each streak corresponding to a KH wave packet convecting downstream along the separated shear layer. The nearly linear slope of these streaks represents a well-defined convection velocity, and the periodic pattern reflects regular vortex shedding. From the temporal spacing between successive streaks, the shedding frequency corresponds to approximately 12 waves per second at this Mach number. This marks the development of a stronger, more coherent shear-layer instability compared to $M_s = 0.15$. At $M_s = 0.25$ in Fig. \ref{fig5}(c), the travelling-wave pattern begins earlier along the suction surface, indicating an upstream shift in the onset of unsteadiness. The KH wave packets display higher slope gradients and more intense signatures of $\omega_y$. The shedding frequency increases to around 15 waves per second, revealing the amplifying effect of compressibility on shear-layer instability. The increased frequency and amplitude lead to quicker breakdown to turbulence, consistent with experimentally observed acceleration of transition with Mach number. At $M_s = 0.30$ and 0.35 in Figs. \ref{fig5}(d) and \ref{fig5}(e), the space–time plot shows dense and closely spaced travelling-wave streaks immediately downstream of the leading edge. The higher frequency of these travelling waves (approximately 16 and 18 waves per second for $M_s = 0.3$ and 0.35, respectively) illustrates a highly unstable and sensitive boundary layer. 

\subsection{Temporal and spatial scale selection}

\begin{figure*}
\centering
\includegraphics[width=.9\textwidth]{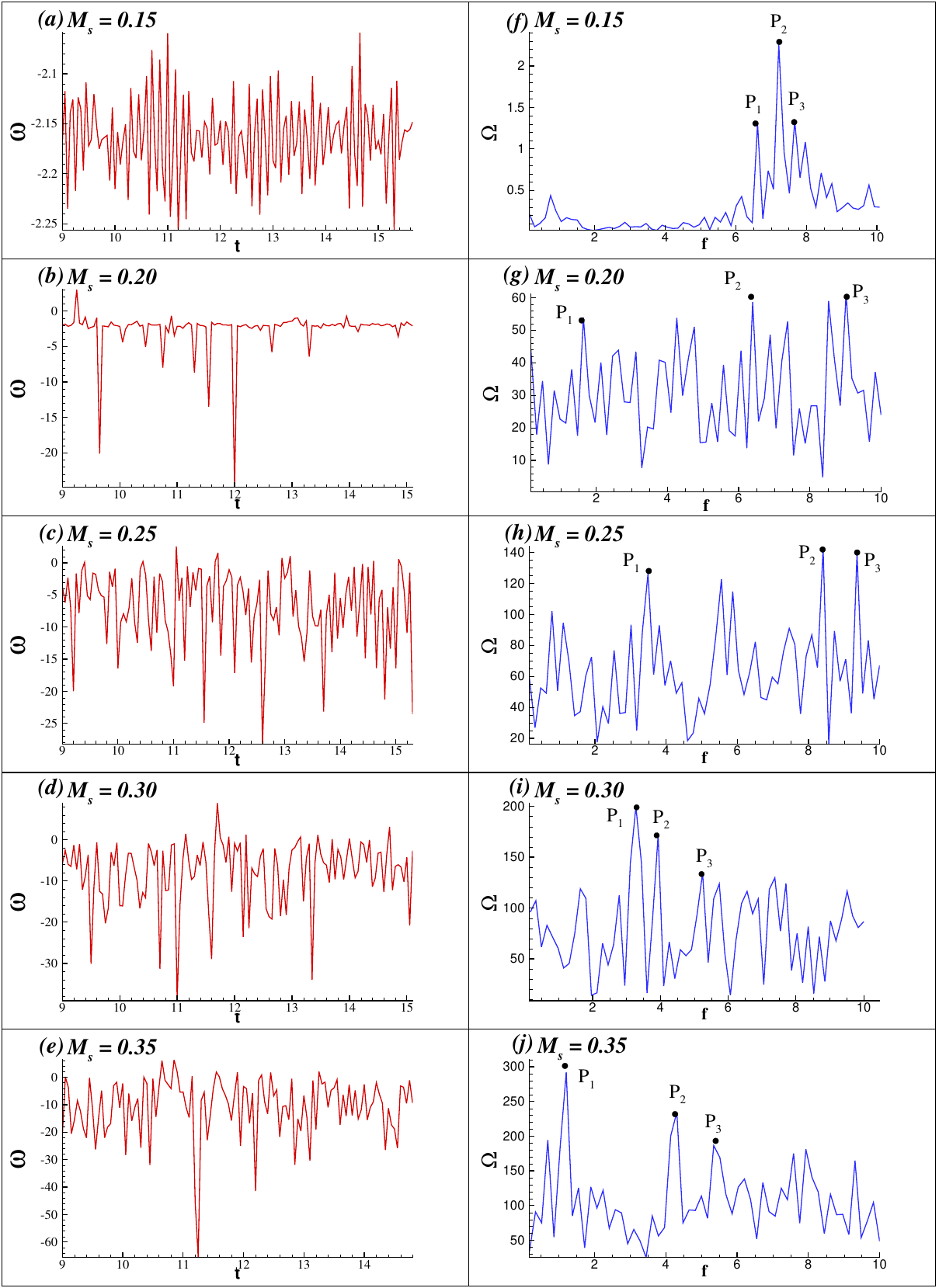}
\caption{Time-series of $\omega_z$ extracted on suction surface of the mid-span plane, for test cases with (a) $M_s = 0.15$, (b) $M_s = 0.20$, (c) $M_s = 0.25$, (d) $M_s = 0.30$ and (e) $M_s = 0.35$. The corresponding spectra in the frequency plane are shown in frames (f), (g), (h), (i) and (j).}
\label{fig6}
\end{figure*}

The time-series of $\omega_z$ and associated spectra are provided at indicated $M_s$ in Fig. \ref{fig6}, for a probe location on the suction surface. The spanwise vorticity signal for $M_s = 0.15$ in Fig. \ref{fig6}(a) remains nearly flat and low-amplitude over the full time window, indicating that the separated shear layer is only weakly unstable and displays minimal periodic activity. Only weak fluctuations appear toward the end of the time interval. The FFT in Fig. \ref{fig6}(f) reflects this with small energy content, three low-amplitude peaks ($P_1$, $P_2$, $P_3$), indicating mild oscillatory behavior rather than organized vortex shedding. This is consistent with a long, laminar separation bubble that remains quiescent until the aft portion of the blade. At $M_s = 0.20$, the time-series in Fig. \ref{fig6}(b) shows a clear onset of periodic oscillations, with the vorticity signal developing a more regular pattern of fluctuations. These correspond to the KH instability waves rolling up the separated shear layer. The FFT spectrum in Fig. \ref{fig6}(g) shows three high amplitude peaks ($P_1$, $P_2$, and $P_3$), compared to $M_s = 0.15$. A dominant frequency is noted at the sub-harmonics of the vortex-shedding frequency, i.e. ∼12 waves per second from the space–time plot in Fig. \ref{fig5}, representing the fundamental shedding mode. 

At $M_s = 0.25$ in Fig. \ref{fig6}(c), the time-series displays stronger, sharper, and higher-amplitude oscillations, indicating a more energetic KH roll-up and earlier transition to nonlinear behaviour. The FFT in Fig. \ref{fig6}(h) shows very strong peaks of comparable magnitude. There is significantly higher energy across the spectrum with an increased dominant frequency. The identified peaks are interharmonics of the identified time scale in the space-time plot in Fig. \ref{fig5}, i.e. ∼15 waves per second. At $M_s = 0.3$, the time-series in Fig. \ref{fig6}(d) becomes more irregular and chaotic, with turbulence superimposed on the periodic signal. This represents a transitional flow where KH waves coexist with turbulent bursts. The FFT in Fig. \ref{fig6}(i) shows a further shift in the amplitude of the dominant frequency peaks. The energy spreads across a broader band with strong secondary peaks reflecting nonlinear interactions and harmonics. The dominant frequency (a sub-harmonic of ~16 waves per second) confirms a faster streak-induced instability at this Mach number. For $M_s = 0.35$ in Fig. \ref{fig6}(e), the vorticity signal is highly energetic and chaotic, indicating a strongly unstable shear layer with frequent turbulent events. The periodic component is still present, but heavily modulated by intermittent bursts. The FFT in Fig. \ref{fig6}(j) reflects very high spectral content, dominant peaks $P_1$-$P_3$ with large amplitudes, a further increase in the fundamental frequency ($\approx 18$ waves per second), and a broad harmonic content due to nonlinear breakdown. For $M_s = 0.25$ and higher $M_s$, streaks induced by the suction-surface separation bubble are long in the streamwise direction. These slowly grow or modulate, and are much larger than the fundamental KH instability wavelength. Because of their large scale and slow evolution, the disturbances these streaks induce naturally have low characteristic frequencies, which is viewed by the dominant frequencies shifting towards the lower range. The exact values of the dominant frequencies ($P_1$-$P_3$) and associated amplitudes are provided in Table \ref{tab2}.

\begin{table}[!ht]
\centering
\caption{Dominant frequencies and their amplitudes marked as $P_1$, $P_2$, $P_3$ in Fig.~6}
\setlength{\tabcolsep}{11pt} 
\begin{tabular}{c c c c c c c}
\hline\hline
$M_{s}$ & $P_1$$(f)$ & $P_1$$(\Omega)$ & $P_2$$(f)$ & $P_2$$(\Omega)$ & $P_3$$(f)$ & $P_3$$(\Omega)$ \\
\hline
0.15 & 6.6170  & 1.3028   & 7.2184  & 2.2579   & 7.6675  & 1.3265 \\
0.20 & 1.2330  & 52.5201  & 6.0143  & 56.1699  & 9.2120  & 60.2278 \\
0.25 & 3.7582  & 124.7625 & 8.5714  & 137.1704 & 9.5007  & 136.7379 \\
0.30 & 3.2048  & 190.9476 & 4.0056  & 161.1875 & 4.8942  & 133.8102 \\
0.35 & 1.8013  & 282.5015 & 4.5923  & 232.5386 & 5.0996  & 185.2913 \\
\hline\hline
\label{tab2}
\end{tabular}
\end{table}

\begin{figure*}
\centering
\includegraphics[width=.9\textwidth]{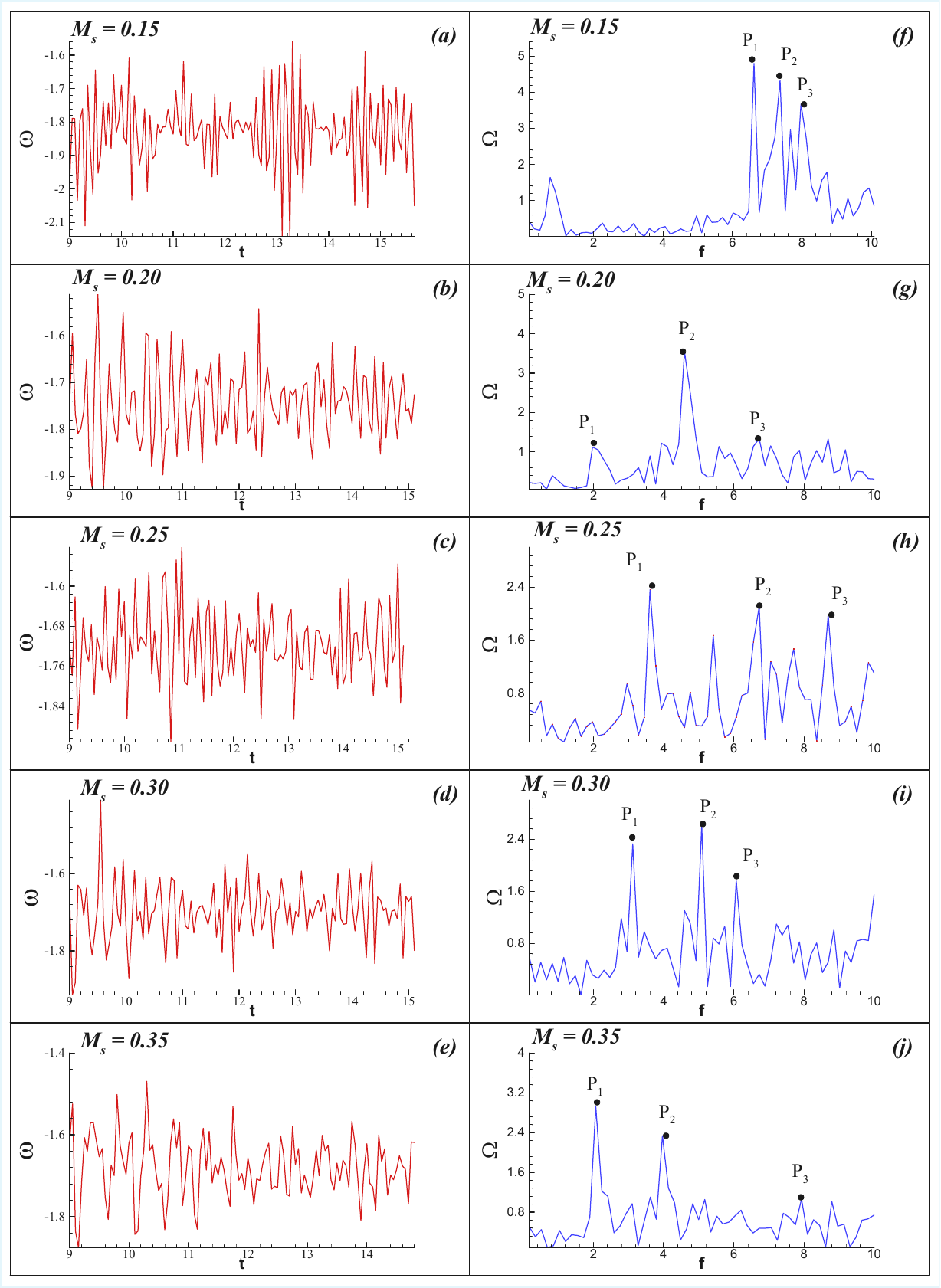}
\caption{Time-series of $\omega_z$ extracted on pressure surface of the mid-span plane, for test cases with (a) $M_s = 0.15$, (c) $M_s = 0.20$, (e) $M_s = 0.25$, (g) $M_s = 0.30$ and (i) $M_s = 0.35$. The corresponding spectra in the frequency plane are shown in frames (b), (d), (f), (h) and (j).}
\label{fig7}
\end{figure*}

In Fig. \ref{fig7}, the time-series of $\omega_z$ and associated spectra are provided for indicated $M_s$, at a probe location on the pressure surface. The dynamics of the vortex shedding pressure-side trailing edge is characterized and its systematic modification with increasing compressibility is explored. At $M_s = 0.15$ in Fig. \ref{fig7}(a), the spanwise vorticity signal exhibits relatively regular oscillations, indicating a weak but coherent vortex shedding process from the pressure-side trailing edge. The FFT spectrum in Fig. \ref{fig7}(b) shows a clearly identifiable dominant peak, $P_1$, accompanied by secondary peaks, $P_2$ and $P_3$, corresponding to harmonics or nonlinear interactions. This is consistent with quasi-periodic vortex shedding from the trailing edge, governed primarily by local geometry and pressure-gradient effects \cite{sieverding1985recent}, with minimal compressibility influence. As the inlet Mach number increases to 0.2 and 0.25 in Figs. \ref{fig7}(c) and \ref{fig7}(e), the time-series displays larger-amplitude and more energetic oscillations, with increased modulation of the signal. This indicates stronger interaction between the pressure-side boundary layer and the trailing-edge wake. In the FFT spectra of Figs. \ref{fig7}(d) and \ref{fig7}(f), the dominant peaks, $P_1$-$P_3$ remain clearly identifiable, but begin to shift toward lower frequencies compared to $M_s = 0.15$. This shift suggests that the characteristic time scale of vortex shedding increases, reflecting a modification of the pressure-side boundary-layer thickness and wake structure with increasing compressibility. The exact shift can be quantified in Table \ref{tab3}, wherein the frequencies $P_1$-$P_3$ shift upstream while the amplitude of the peaks remain comparable. This is in contrast with the suction surface values in Table \ref{tab2}, where the Fourier amplitudes increase significantly as a function of $M_s$.

At higher $M_s = 0.3$ and 0.35 in Figs. \ref{fig7}(g) and \ref{fig7}(i), the spanwise vorticity signal becomes more irregular and strongly amplitude-modulated which is indicative of unsteady, large-scale vortex shedding interspersed with intermittent burst-like events. The FFT spectra in Figs. \ref{fig7}(h) and \ref{fig7}(j) reveal a pronounced shift of all dominant peaks $P_1$-$P_3$ toward lower frequency values, accompanied by a more broadband distribution over a wider range of frequencies. This behavior implies a reduction in shedding frequency, dominance of larger-scale vortical structures in the pressure-side wake, enhanced interaction between the pressure-side boundary layer, trailing-edge separation, and the global wake \cite{michelassi2015compressible}.

\begin{table}[!ht]
\centering
\caption{Dominant frequencies and their amplitudes marked as $P_1$, $P_2$, $P_3$ in Fig.~7}
\setlength{\tabcolsep}{11pt} 
\begin{tabular}{c c c c c c c}
\hline\hline
$M_{s}$ & $P_1$$(f)$ & $P_1$$(\Omega)$ & $P_2$$(f)$ & $P_2$$(\Omega)$ & $P_3$$(f)$ & $P_3$$(\Omega)$ \\
\hline
0.15 & 6.6127  & 4.7885  & 7.3679  & 4.3264  & 7.9716  & 3.6228 \\
0.20 & 1.6665  & 1.1135  & 4.6879  & 3.5198  & 6.7221  & 1.3304 \\
0.25 & 3.5063  & 2.3611  & 6.7207  & 2.1059  & 8.6884  & 1.9775 \\
0.30 & 3.3148  & 2.3285  & 5.0819  & 2.6174  & 6.0654   & 1.7679 \\
0.35 & 2.0689  & 2.9247  & 3.9656  & 2.3366  & 7.9309   & 1.0608  \\
\hline\hline
\label{tab3}
\end{tabular}
\end{table}

\begin{figure*}
\centering
\includegraphics[width=.9\textwidth]{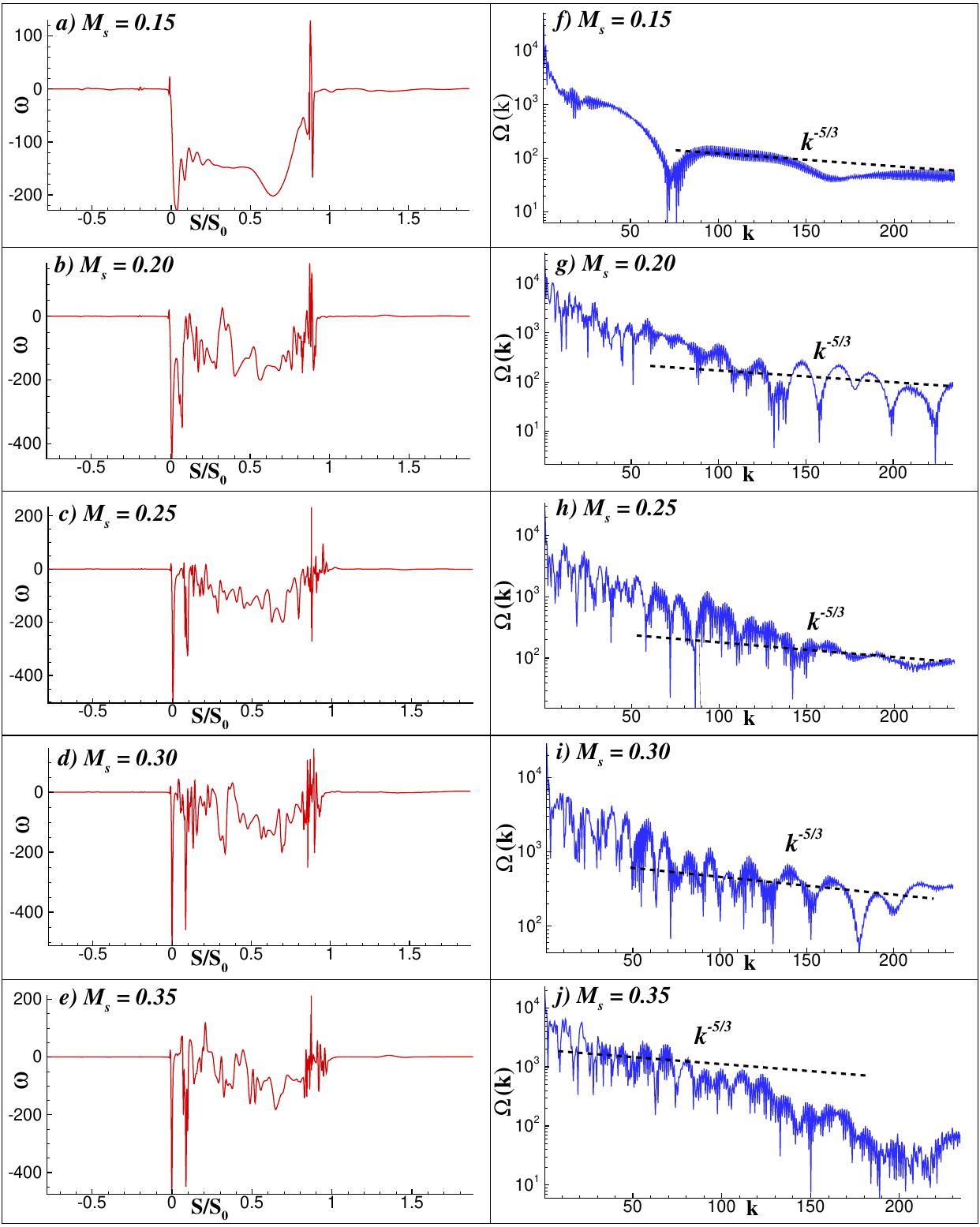}
\caption{The streamwise distribution of $\omega_z$ extracted near suction surface of the T106A LPT blade from the mid-span plane, for test cases with (a) $M_s = 0.15$, (b) $M_s = 0.20$, (c) $M_s = 0.25$, (d) $M_s = 0.30$ and (e) $M_s = 0.35$. The corresponding spectra in the wavenumber plane are shown in frames (f)-(j), along with the $k^{-5/3}$ scaling line.}
\label{fig8}
\end{figure*}

In Fig. \ref{fig8}, the streamwise distribution of $\omega_z$ is extracted at a near-wall location on the suction surface in the mid-span plane for various $M_s$. The corresponding FFT are shown as a function of the wavenumber, $k$. Across all cases, the spectra exhibit a clear $k^{-5/3}$ region, indicative of a Kolmogorov-like inertial range and confirming the presence of turbulent energy cascades in the near-wall flow \cite{tennekes1972first}. A key observation is that the location of the $k^{-5/3}$ scaling shifts systematically with $M_s$, revealing how dominant spatial scales of turbulence change with compressibility. For $M_s = 0.15$ and 0.2 shown in Figs. \ref{fig8}(a) and \ref{fig8}(b), the $k^{-5/3}$ scaling appears only at relatively high wavenumbers, i.e. at small spatial scales. This indicates that flow over the suction surface remains largely laminar or weakly transitional over much of the chord. Turbulence is generated late in the transition process, typically downstream of the separation bubble and energy is injected initially at small scales through localized shear-layer roll-up and secondary instabilities. The inertial cascade therefore develops only after sufficient breakdown, populating the higher $k$-range. In physical terms, the near-wall turbulence at low $M_s$ is fine-scaled and intermittent, consistent with delayed transition and weaker compressibility effects. At $M_s = 0.25$ shown in Fig. \ref{fig8}(c), the $k^{-5/3}$ region extends toward moderate wavenumbers, indicating a broadening of dynamically active spatial scales. This reflects earlier onset of separation-induced transition, and stronger three-dimensionality due to enhanced streak–shear-layer interaction \cite{duan2023effects}. Energy injection occurs at larger spatial scales than at lower $M_s$, followed by a more developed cascade. The near-wall flow transitions from being dominated by small-scale breakdown to a multi-scale turbulent structure. For $M_s = 0.3$ and 0.35 shown in Figs. \ref{fig8}(d) and \ref{fig8}(e), the $k^{-5/3}$ scaling is observed at much lower wavenumbers, corresponding to larger spatial scales. This shift signifies that transition is triggered closer to the leading edge. Large-scale structures such as Klebanoff streaks, spanwise waviness, and global separation-bubble modes dominate the flow. Compressibility enhances receptivity and promotes energy injection at large scales, rather than through gradual small-scale breakdown \cite{lele1994compressibility}. The turbulent cascade therefore starts from low $k$ and extends downscale. Physically, this indicates a transition pathway closer to bypass-type transition, where large coherent structures directly feed the turbulent cascade.

\subsection{Boundary layer characteristics on suction surface}

In this section, we examine the evolution of the suction-surface boundary layer under varying $M_s$, with particular emphasis on the effects of separation-induced transition and compressibility. Key streamwise variation of boundary-layer quantities - including time-averaged free-stream velocity at the boundary-layer edge ($U_{fs}$), momentum thickness ($\theta_m$), skin-friction coefficient ($C_f$), and coefficient of static pressure ($C_p$) - are analyzed. These metrics provide a comprehensive description of boundary-layer growth, separation and reattachment behavior, and onset and progression of transition. 

\begin{figure*}
\centering
\includegraphics[width=.8\textwidth]{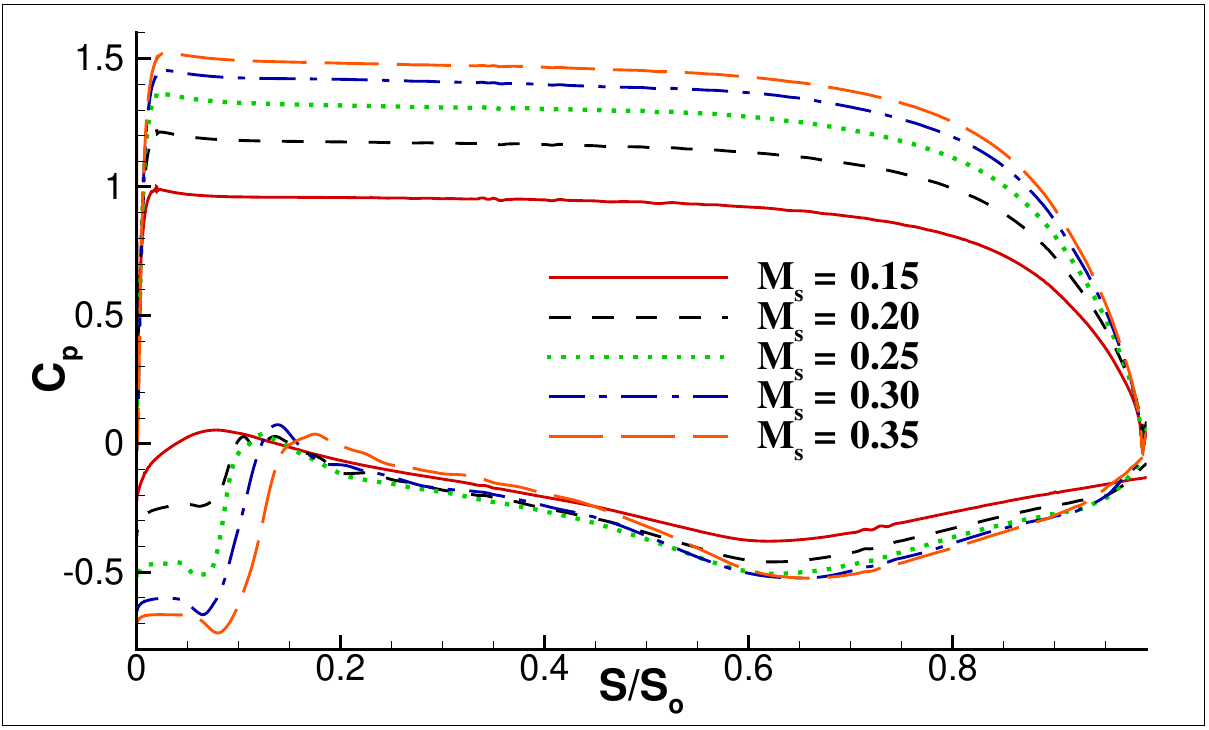}
\caption{Streamwise variation of time-averaged coefficient of pressure, $C_p$ for test cases with $M_s = 0.15$, 0.20, 0.25, 0.30 and 0.35.}
\label{fig9}
\end{figure*}

Figure \ref{fig9} compares the time-averaged distribution of $C_p$, along the suction surface for different $M_s$. Two prominent features are observed consistently across all cases: a leading-edge plateau and a trailing-edge plateau, both of which are characteristic signatures of separation bubbles in LPT flows \cite{fiore2023t106}. Near aft portion of the suction surface, a pronounced plateau in $C_p$ indicates presence of a long trailing-edge separation bubble. Such an extended region of nearly constant pressure arises when boundary layer encounters a strong adverse pressure gradient and separates, resulting in a shear layer that shields the wall from pressure recovery \cite{garai2015dns}. The trailing-edge separation bubble is of particular concern in turbine design because it increases aerodynamic losses, thickens the wake, and degrades stage efficiency \cite{denton1993loss}. Among the cases examined, the longest plateau (separation bubble) occurs at $M_s = 0.15$. As $M_s$ increases, extent of this plateau progressively reduces, indicating a shorter and weaker separation bubble. A second plateau is visible near the leading edge of the suction surface, signifying the presence of a leading-edge separation bubble. This bubble forms due to rapid acceleration around the blade nose followed by a localized adverse pressure gradient. As $M_s$ increases, the streamwise extent of this leading-edge plateau decreases, indicating a reduction in the size of the leading-edge separation bubble. Higher $M_s$ accelerates transition within the separated shear layer, leading to enhanced momentum exchange and earlier reattachment, thereby shortening the bubble \cite{sengupta2023compressibility}. 

Under the imposed inflow condition with an incidence angle of $45^{\circ}$, a stagnation point forms on the pressure surface rather than exactly at the leading edge. The incoming flow must therefore be redirected around the blade nose to enter the suction surface. This redirection generates a strong adverse pressure gradient locally, and when sufficiently strong, it leads to flow reversal within the near-wall region of the boundary layer \cite{ranjan2014direct}. Such flow reversal is observed for all $M_s$, indicating that the leading-edge separation is primarily driven by blade geometry and incidence rather than compressibility alone. A notable trend emerges when comparing the overall $C_p$ distributions, i.e. increasing $M_s$ has an effect analogous to increasing free-stream disturbances. In both cases separation is delayed, bubbles are shorter in the time-averaged sense, and reattachment occurs earlier. Higher $M_s$ enhances unsteady shear-layer activity and promotes earlier transition through compressibility-driven instability mechanisms. This mirrors the role of free-stream turbulence, which also accelerates transition and strengthens near-wall momentum transport \cite{sengupta2020effects, sengupta2024separation}.

\begin{figure*}
\centering
\includegraphics[width=.8\textwidth]{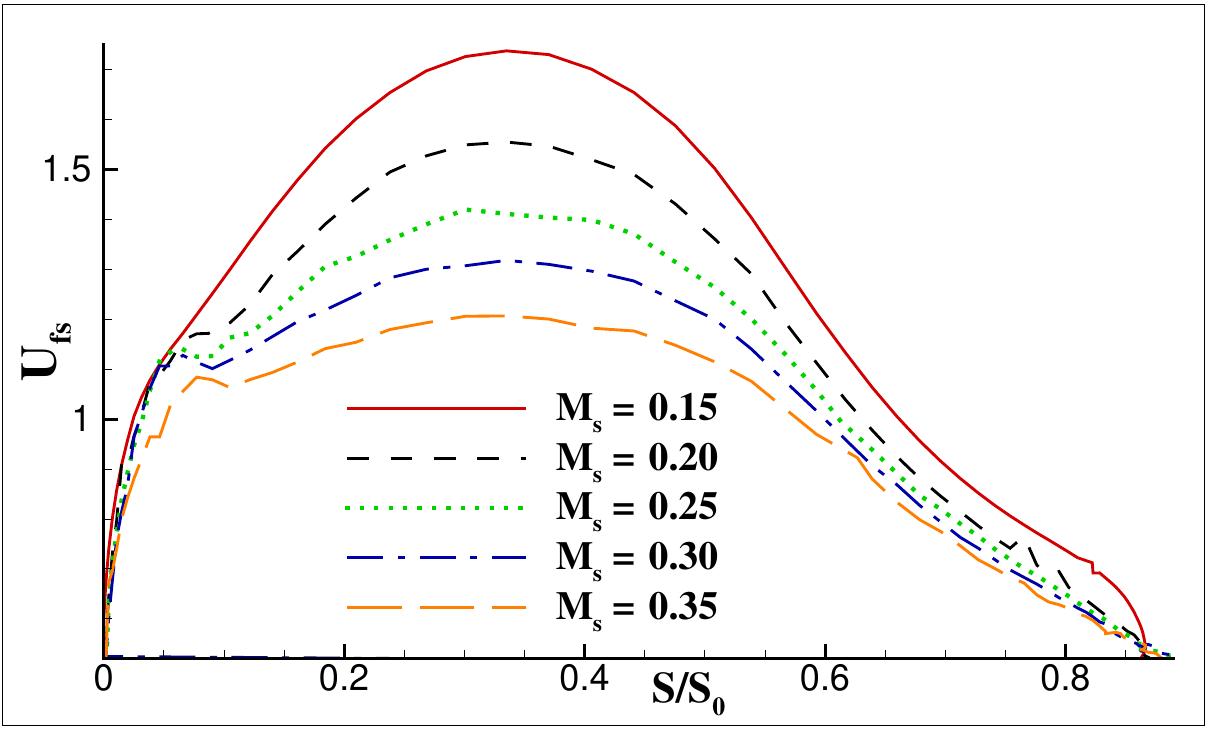}
\caption{Streamwise variation of free-stream velocity, $U_{fs}$ at the edge of the boundary layer for test cases with $M_s = 0.15$, 0.20, 0.25, 0.30 and 0.35. }
\label{fig10}
\end{figure*}

Figure \ref{fig10} shows the streamwise variation in the time-averaged free-stream velocity, $U_{fs}$ at the edge of the boundary layer, along the suction surface for the indicated $M_s$. For all values of $M_s$, the suction-side boundary layer initially experiences a strong favorable pressure gradient, causing acceleration of the external flow until the peak suction location at $S/S_0 = 0.42$. The edge velocity reaches a maximum value of approximately 1.8 for $M_s = 0.15$. Beyond this point, the pressure gradient becomes adverse, and the boundary layer progressively loses momentum. At lower Mach numbers ($M_s = 0.15$), the boundary layer remains largely laminar when it encounters the adverse pressure gradient. Owing to its low near-wall momentum, it separates readily, forming a long separation bubble \cite{garai2015dns, michelassi2015compressible}. The separated shear layer shields the wall from the external flow, resulting in a nearly stationary region beneath it, which manifests as an extended velocity plateau in the distribution. Reattachment occurs slowly because the separated shear layer is only weakly unstable, leading to a gradual deceleration downstream. As $M_s$ increases, several coupled mechanisms act to shorten the plateau and steepen the post-separation deceleration. With increased compressibility, the receptivity of the separated shear layer to instability waves increases. Further, stronger Kelvin–Helmholtz-type roll-ups and earlier three-dimensional breakdown accelerate transition within the shear layer \cite{lele1994compressibility}. Once transition occurs, turbulent fluctuations efficiently transport high-momentum fluid from the outer flow toward the near-wall region. This enhanced momentum exchange allows the boundary layer to better resist the adverse pressure gradient, reducing the duration for which the flow remains separated. Due to the higher near-wall momentum, reattachment occurs closer to the separation point at higher $M_s$. As a result, the streamwise extent of the nearly constant-velocity plateau beneath the separated shear layer decreases.

\begin{figure*}
\centering
\includegraphics[width=.8\textwidth]{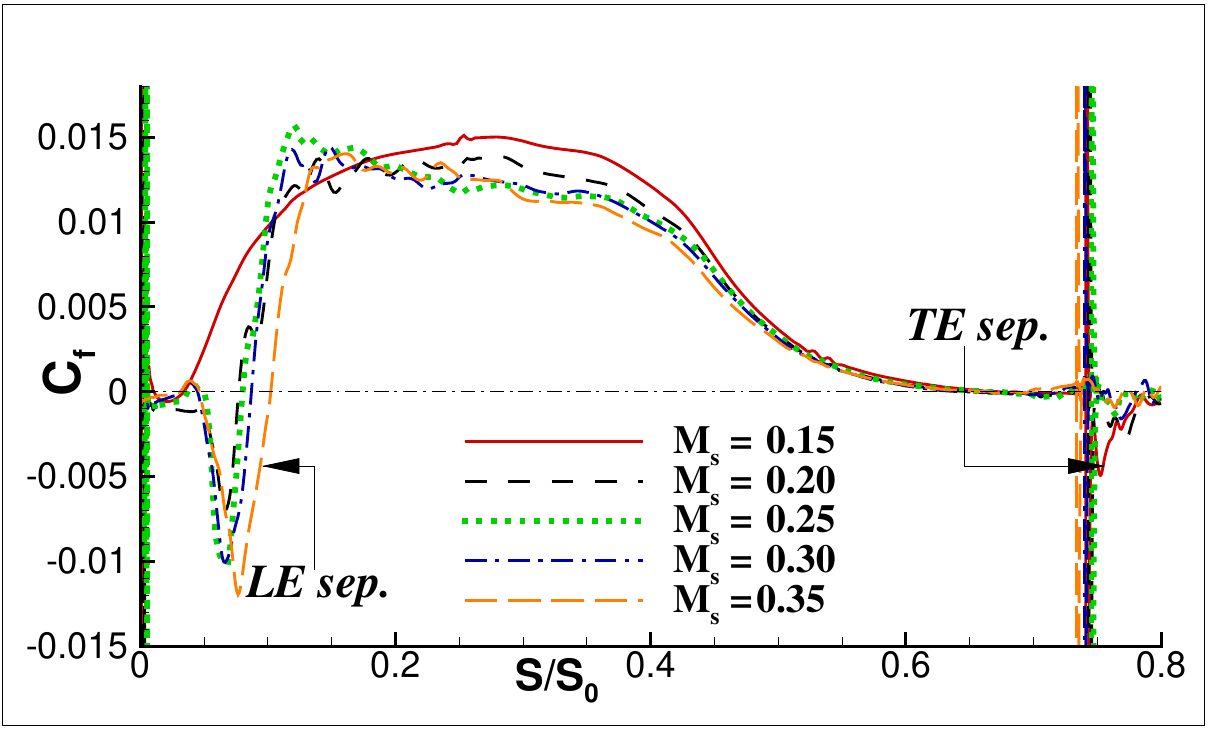}
\caption{Streamwise variation of time-averaged skin friction coefficient, $C_f$ for test cases with $M_s = 0.15$, 0.20, 0.25, 0.30 and 0.35.}
\label{fig11}
\end{figure*}

Figure \ref{fig11} presents streamwise variation of the time-averaged skin-friction coefficient, $C_f$ along the suction surface for different values of $M_s$, as indicated. The $C_f$ distribution provides an indication of near-wall shear, enabling identification of separation, reattachment, transition, and relaminarization processes \cite{schlichting1961boundary}. For all $M_s$, a leading-edge separation bubble \cite{simoni2015off, wissink2006influence} is clearly identified by a region of negative $C_f$ near $S/S_0 \approx 0.05$. This bubble forms due to the strong local adverse pressure gradient generated by flow turning around the blade nose at high incidence. Notably, earlier transition within the leading-edge bubble is observed for $M_s = 0.15$, as separated shear layer is weakly unstable and supports amplified instability waves that trigger transition locally within the bubble. As $M_s$ increases, transition within this bubble occurs further downstream, but the overall streamwise extent of the separated region decreases. Immediately downstream of the leading-edge bubble, a transition region is evident from the rapid rise of $C_f$ from negative to positive values. This region corresponds to the breakdown of the separated shear layer and the onset of turbulent mixing. The transition process occurs earlier and more abruptly with increasing $M_s$, consistent with compressibility-enhanced instability growth. Following reattachment, a plateau in $C_f$ is observed, indicative of relaminarization of the boundary layer under a favorable pressure gradient. This plateau signifies reduced near-wall turbulence production and stabilization of the boundary layer. The relaminarization region persists longest for $M_s = 0.35$, although the difference across $M_s$ is marginal. At higher $M_s$, stronger acceleration and enhanced mixing during reattachment lead to a fuller velocity profile that can sustain a quasi-laminar state over longer distance before encountering the next adverse pressure gradient. A distinct dip in $C_f$ is observed near peak suction location ($S/S_0 \approx 0.42$), corresponding to onset of a strong adverse pressure gradient. The reduced wall shear reflects boundary-layer deceleration and loss of near-wall momentum, preconditioning of boundary layer for aft-chord separation. Near aft portion of blade, another region of negative $C_f$ indicates presence of a trailing-edge separation bubble. The size and strength of this bubble decrease with increasing $M_s$. Downstream of trailing-edge bubble, a partial re-transition is observed, marked by recovery of positive $C_f$. This indicates renewed turbulence production as the flow enters the wake region. Based on time-averaged $C_f$-distribution, flow has been classified into following regions: (i) leading edge separation bubble near $S/S_0 = 0.05$, (ii) transition region persisting from $S/S_0 = 0.01$ to 0.1, (iii) relaminarization of flow from $S/S_0 = 0.1$ to 0.3, (iv) adverse pressure gradient near $S/S_0 = 0.42$, (v) separation bubble near aft portion of blade at approximately $S/S_0 = 0.75$, and (vi) eventual re-transition of flow beyond $S/S_0 = 0.75$. 

In a low-pressure turbine (LPT) cascade, the total pressure loss downstream of a blade row serves as a key metric for quantifying the aerodynamic losses incurred by the blade. This loss can be expressed using the momentum-based formulation proposed by Denton \cite{denton1993loss} as

\begin{equation}
\zeta = \biggl ( \frac{-C_{pb} t_{TE}}{p_b cos\alpha_2} \biggr ) + \biggl ( \frac{ \delta_{TE} + t_{TE}}{p_b cos\alpha_2} \biggr )^2 + \biggl ( \frac{2 \theta_{TE}}{p_b cos\alpha_2} \biggr )
\label{mom_loss}
\end{equation}

\noindent where $C_{pb}$ denotes the base pressure coefficient, $t_{TE}$ is the trailing-edge thickness, $\alpha_2$ is the exit flow angle measured relative to the axial direction, and $p_b$ represents the blade pitch. The displacement thickness and momentum thickness at the trailing edge are denoted by $\delta_{TE}$ and $\theta_{TE}$, respectively. The first term on the right-hand side of Eq. \eqref{mom_loss} accounts for the loss associated with the low base pressure acting on the trailing edge. Previous studies have shown that this contribution is relatively small, typically on the order of 3\% of the total loss \cite{curtis1997development}. The second term represents losses arising from blockage effects due to the combined influence of trailing-edge thickness and boundary-layer displacement thickness, contributing approximately 7\% to the overall loss. The dominant contribution, accounting for nearly 90\% of the total loss, originates from the mixed-out loss of the boundary layers, represented by the third term. This component is effectively proportional to the trailing-edge momentum thickness, underscoring the critical role of boundary-layer development in determining LPT blade losses \cite{coull2012predicting}.

\begin{figure*}
\centering
\includegraphics[width=.8\textwidth]{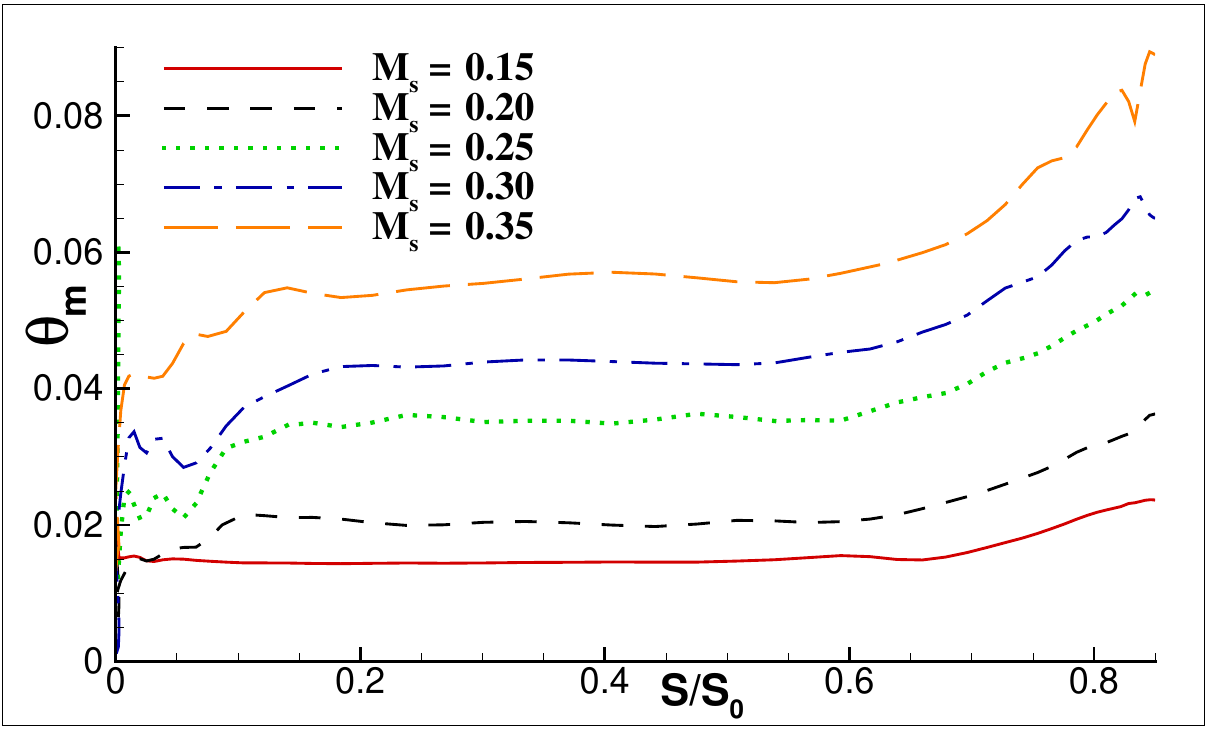}
\caption{Streamwise variation of time-averaged momentum thickness, $\theta_m$ on the suction surface of T106A LPT cascade for test cases with $M_s = 0.15$, 0.20, 0.25, 0.30 and 0.35.}
\label{fig12}
\end{figure*}

Figure \ref{fig12} shows the streamwise variation of time-averaged momentum thickness, $\theta_{m}$, on suction surface normalized with respect to surface length, $S_0$. Increasing $M_s$ from 0.15 to 0.35, one observes a rise in the momentum thickness. Although a smaller streamwise extent of separation and a delayed separation are reported for higher $M_s$ in Fig. \ref{fig11}, the overall height of individual separation bubbles (in a time-averaged sense increases with $M_s$. There is a marked departure of $\theta_m$ at trailing edge, suggesting that compressibility has a major role in determining blade profile losses. The momentum thickness at the trailing edge increases from 0.02 to 0.09 for $M_s$ increasing from 0.15 to 0.35, i.e. a $350\%$ increase in $\theta_{TE}$. The momentum thickness, represents the cumulative loss of streamwise momentum within the boundary layer relative to the external flow. It depends not only on whether the flow is separated, but also on shape and fullness of the velocity profile, intensity of turbulent mixing, and strength of unsteady shear-layer dynamics. Skin-friction profile of Fig. \ref{fig11} shows that with increasing $M_s$, both leading-edge and trailing-edge separation bubbles are delayed, occupy a smaller streamwise extent, and reattach earlier due to enhanced transition and mixing. However, earlier transition within the separated shear layer at higher $M_s$ leads to stronger velocity fluctuations, increased turbulent shear stresses, and enhanced wall-normal momentum transport. This results in a thicker, more energetic boundary layer downstream of reattachment, even though the separated region itself is shorter. Although separation bubbles become shorter with increasing $M_s$, their time-averaged vertical extent (bubble height) increases, as is evident from Figs. \ref{fig3} and \ref{fig4}. This has two important consequences: (i) separated shear layer is lifted farther away from the wall, and (ii) near-wall region experiences a stronger momentum deficit during separation and early reattachment. When integrated across the boundary layer thickness, this leads to a larger momentum deficit, and hence a higher $\theta_m$, despite reduced separation length. Momentum thickness is cumulative along the blade surface. By the time the flow reaches the trailing edge, the boundary layer at higher $M_s$ has undergone multiple transition–reattachment cycles, turbulent mixing has acted over a longer portion of the suction surface, and the wake deficit becomes significantly larger. This explains the dramatic rise in trailing-edge momentum thickness with $M_s$.

\subsection{Budget of turbulent kinetic energy and compressible enstrophy}

Turbulent kinetic energy (TKE), commonly denoted by $k$, is a measure of the kinetic energy contained in the velocity fluctuations of a turbulent flow. It quantifies how energetic the turbulence is and is defined as $k = \frac{1}{2} \overline{u'^2} + \overline{v'^2} + \overline{w'^2}$, where $u'$, $v'$, and $w'$ are the fluctuating components of velocity in the streamwise, wall-normal, and spanwise directions, respectively, and the overbar denotes time averaging. Physically, TKE represents the intensity of turbulence and the amount of energy available in eddies to transport momentum, heat, and mass \cite{alam2000direct}. In separation-induced transition, the boundary layer separates while still laminar, forming a separated shear layer. Transition typically occurs when instabilities within this shear layer amplify and break down to turbulence. Thus, a rapid rise in TKE marks the onset of transition inside or just downstream of the separation bubble. It serves as a clear diagnostic for identifying where and how transition occurs along the blade.

\begin{figure*}
\centering
\includegraphics[width=.8\textwidth]{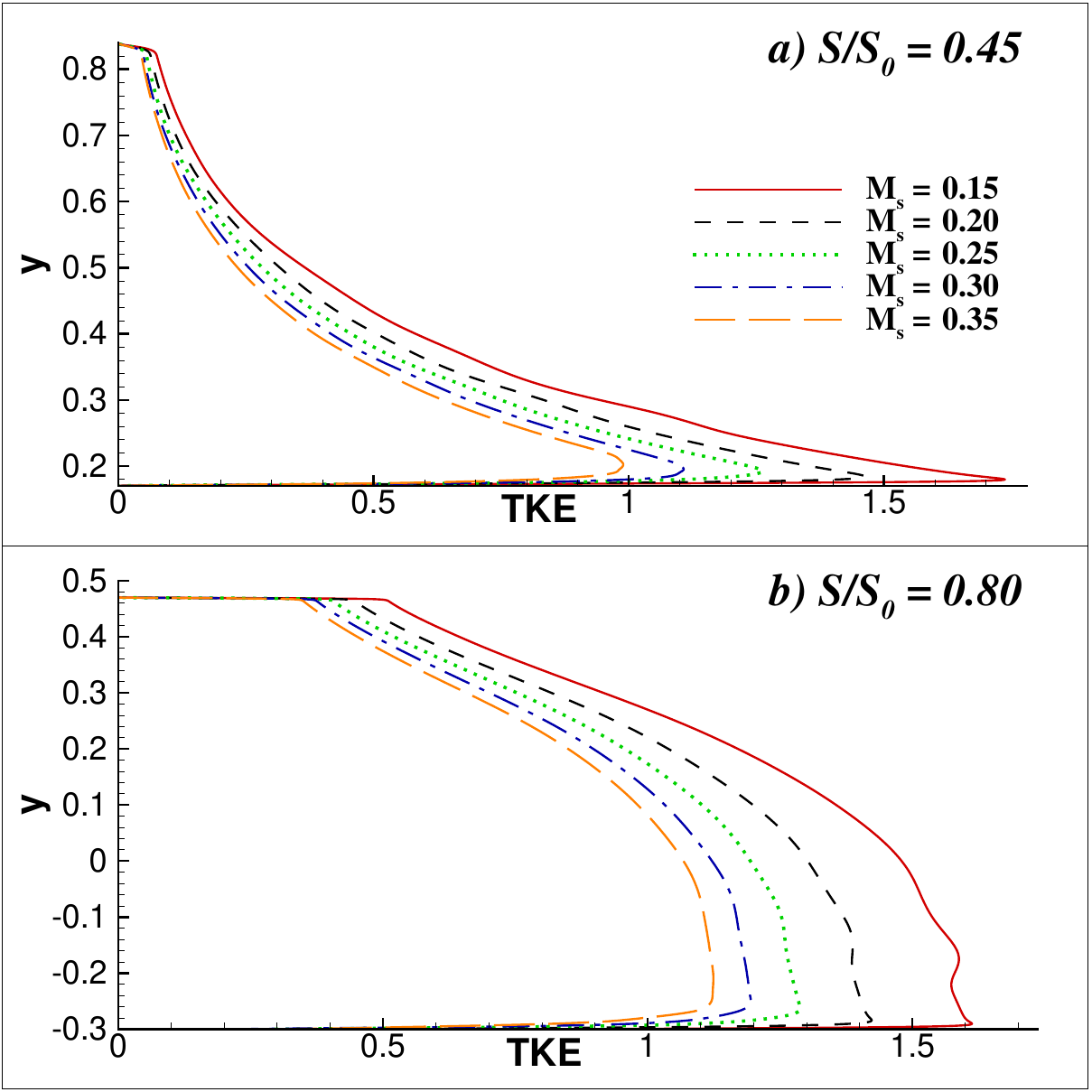}
\caption{Production term of turbulent kinetic energy (TKE) at indicated streamwise locations, $(a) S/S_0 = 0.40$ and $(b) S/S_0 = 0.80$ for test cases with $M_s = 0.15$, 0.20, 0.25, 0.30 and 0.35.}
\label{fig13}
\end{figure*}

Figure \ref{fig13} presents the wall-normal variation of time-averaged TKE, normalized by the surface length, $S_0$, at two streamwise locations along the suction surface: $S/S_0 = 0.45$, located downstream of the peak adverse pressure gradient, and $S/S_0 = 0.80$, near the trailing edge. The distributions are shown for $M_s$ ranging from 0.15 to 0.35. At both streamwise locations, the maximum TKE is observed for the lowest Mach number, $M_s = 0.15$, with TKE levels progressively decreasing as $M_s$ increases. At lower $M_s$, the separated shear layer remains laminar over a longer distance before undergoing transition. When breakdown occurs, it is characterized by strong Kelvin–Helmholtz roll-ups, vigorous vortex pairing, and intense three-dimensional breakdown. These processes generate large-amplitude velocity fluctuations, resulting in higher TKE levels. In contrast, at higher $M_s$, transition is triggered earlier via streak-induced breakdown. This leads to less energetic, and more distributed turbulence production. Thus, there is reduced peak TKE despite increased overall momentum loss. Earlier transition does not necessarily imply higher local TKE, particularly when the fluctuating component of energy is spread over a broader region. The peak TKE occurs closer to the wall for lower $M_s$, whereas it shifts away from the wall as $M_s$ increases. At lower $M_s$, separated shear layer lies closer to the wall and remains coherent over a longer streamwise distance, concentrating turbulence production in the near-wall region. As $M_s$ increases, enhanced streak-induced instability and earlier reattachment lift the shear layer away from the wall, relocating the zone of maximum turbulence production farther into the outer boundary layer. The TKE levels at $S/S_0 = 0.45$ are consistently higher than those at $S/S_0 = 0.8$ for all $M_s$ shown here. The location in Fig. \ref{fig13}(a) lies immediately downstream of the peak adverse pressure gradient and coincides with active shear-layer instability, breakdown of the separation bubble, and peak turbulence production. Near the trailing edge shown in Fig. \ref{fig13}(b), the flow has already undergone significant mixing and partial relaminarization. As a result, turbulence production decreases and TKE decays due to viscous dissipation and redistribution of energy into the wake. For the lower Mach numbers ($M_s = 0.15$ and 0.2), secondary peaks appear in the wall-normal TKE distribution near the trailing edge. These secondary maxima arise from intermittent re-transition events associated with the trailing-edge separation bubble, shear-layer roll-up near the aft portion of the blade, and interaction between the reattached boundary layer and the developing wake. At higher $M_s$, these secondary peaks are absent or significantly weakened as trailing-edge separation bubble is shorter and less coherent, transition occurs earlier upstream, and flow entering the trailing-edge region is already highly mixed, suppressing localized TKE amplification. 

Enstrophy is a measure of the intensity of vorticity in a flow and is defined as, $\Omega = \vec{\omega}\cdot \vec{\omega}$. It quantifies how strongly the flow is rotating at small scales. In turbulence, especially near walls and in shear layers, enstrophy is closely tied to shear-layer roll-up, vortex stretching and tilting, and dissipation of kinetic energy. In incompressible flows, enstrophy production and destruction provide a clear picture of how turbulence is generated, amplified, and dissipated \cite{doering1995applied}. Separation-induced transition in low-pressure turbines is governed by laminar separation bubbles, Kelvin–Helmholtz instabilities, and breakdown into three-dimensional vortical structures. All these processes are vorticity-dominated, not energy-dominated. Thus, enstrophy is a more direct indicator than TKE for onset of shear-layer instability, vortex roll-up and pairing, transition and reattachment dynamics, and near-wall turbulence generation. On the suction surface of an LPT, TKE determines how energetic the turbulence is, while enstrophy tells where and how vortices are being created. The classical (incompressible) enstrophy transport equation \cite{sengupta2018enstrophy} assumes constant density, zero velocity divergence with no dilatational effects. These assumptions do not hold in compressible LPT flows, even at moderate Mach numbers ($M_s \approx 0.15-0.35$), as density varies due to acceleration and pressure gradients, velocity divergence is non-zero, dilatational and baroclinic effects become important, and pressure–density misalignment generates vorticity. Thus, incompressible enstrophy fails to capture key physics in compressible separation-induced transition. The transport equation of compressible enstrophy \cite{suman2022novel} serves as a valuable tool for analyzing the generation, distribution, and evolution of enstrophy during transition to turbulence in various internal and external flows. 

The derivation of the compressible enstrophy transport equation (CETE) from the compressible Navier–Stokes equations was first presented in an earlier work \cite{suman2022novel}. The CETE offers a rigorous framework for quantifying vorticity dynamics by accounting for multiple compressibility-related mechanisms \cite{sengupta2025bifurcation, joshi2025comparing}. Specifically, it captures vorticity generation due to both velocity shear and density gradients (baroclinic effects), the amplification or attenuation of enstrophy arising from flow dilatation, and the coupling between rotation-induced pressure–density fields and vorticity evolution within boundary layers and wakes. The application of the CETE to hydrodynamic instabilities has revealed that, in buoyancy-dominated flows, viscous terms play a leading role in governing enstrophy evolution. In contrast, for advection-dominated flows, vortex stretching emerges as a major contributor alongside viscous effects. In many cases, enstrophy amplification can be traced to baroclinic contributions, particularly after the emergence of quasi-periodic coherent vortical structures in the flow. For external aerodynamic flows \cite{sengupta2025compressible, sengupta2024thermal}, however, the dominant contributions to enstrophy arise from vortex stretching and viscous stress terms, highlighting a distinct balance of mechanisms compared to buoyancy-driven systems. The individual terms comprising the CETE \cite{suman2022novel} are summarized below:

	\begin{equation}
		\begin{aligned}
			\frac{D{\Omega }}{Dt}
			= & \; 2\vec{\omega } \cdot \left[(\vec{\omega} \cdot {\nabla}) \vec{V}\right] - 2({\nabla} \cdot \vec{V}) \Omega \\
			& + \left(\frac{2}{\rho^{2}}\right) \vec{\omega } \cdot \left[\left({\nabla \rho} \times {{\nabla p}}\right)\right] -\left(\frac{2}{\rho^{2}}\right)\vec{\omega } \cdot \left[{\nabla \rho} \times {\nabla} \left(\lambda ({\nabla} \cdot \vec{V})\right)\right] \\
			& +\left(\frac{4}{\rho}\right)\vec{\omega } \cdot \left[{\nabla} \times \left[{\nabla} \cdot \left(\mu S \right)\right]\right] -\left(\frac{4}{\rho^2}\right)\vec{\omega } \cdot \left({\nabla \rho} \times ({\nabla} \cdot \left(\mu S\right)) \right)
		\end{aligned}
		\label{CETE}
	\end{equation}
	
The various terms of Eq. \eqref{CETE} are as follows:

\begin{itemize}
		
\item $2\vec{\omega } \cdot \left[(\vec{\omega} \cdot {\nabla}) \vec{V}\right]$ : Contribution to enstrophy due to vortex stretching (T1). 
		
\item $({\nabla} \cdot \vec{V}) \Omega$: Enstrophy growth/decay due to compressibility (T2). 
		
\item $\left(\frac{1}{\rho^{2}}\right) \vec{\omega } \cdot \left[\left({\nabla \rho} \times {{\nabla p}}\right)\right]$: Contribution from baroclinic term due to misalignment of gradients of pressure and density (T3). 
		
\item $\left(\frac{1}{\rho^{2}}\right)\vec{\omega } \cdot \left[{\nabla \rho} \times {\nabla} \left(\lambda ({\nabla} \cdot \vec{V})\right)\right]$ : Contribution due to misalignment of vorticity and bulk viscosity gradients (T4).
		
\item $\left(\frac{1}{\rho}\right)\vec{\omega } \cdot \left[{\nabla} \times \left[{\nabla} \cdot \left(\mu S \right)\right]\right]$: Diffusion of enstrophy due to viscous action (T5).
		
\item $\left(\frac{1}{\rho^2}\right)\vec{\omega } \cdot \left({\nabla \rho} \times ({\nabla} \cdot \left(\mu S\right)) \right)$: Contribution due to misalignment of gradients of density and divergence of viscous stresses (T6).
\end{itemize}

Vortex stretching term (T1) is dominant in shear layers and near reattachment locations. It controls amplification of three-dimensional vortices in the separated shear layer \cite{lin2022physical}. Dilatation terms, on the other hand, are unique to compressible flows. These either amplify or attenuate enstrophy depending on local expansion/compression, and becomes significant near peak suction and reattachment zones \cite{sarkar1991analysis}. Baroclinic torque is generated when pressure and density gradients are misaligned \cite{jahanbakhshi2015baroclinic}. It is particularly important near leading-edge acceleration, regions where separation bubbles are formed, and regions with adverse pressure gradients. Baroclinic torque directly contributes to vorticity, and is completely absent in incompressible theory. Viscous dissipation terms represent destruction of enstrophy at small scales and are closely linked to turbulent kinetic energy dissipation.

\begin{figure*}
\centering
\includegraphics[width=\textwidth]{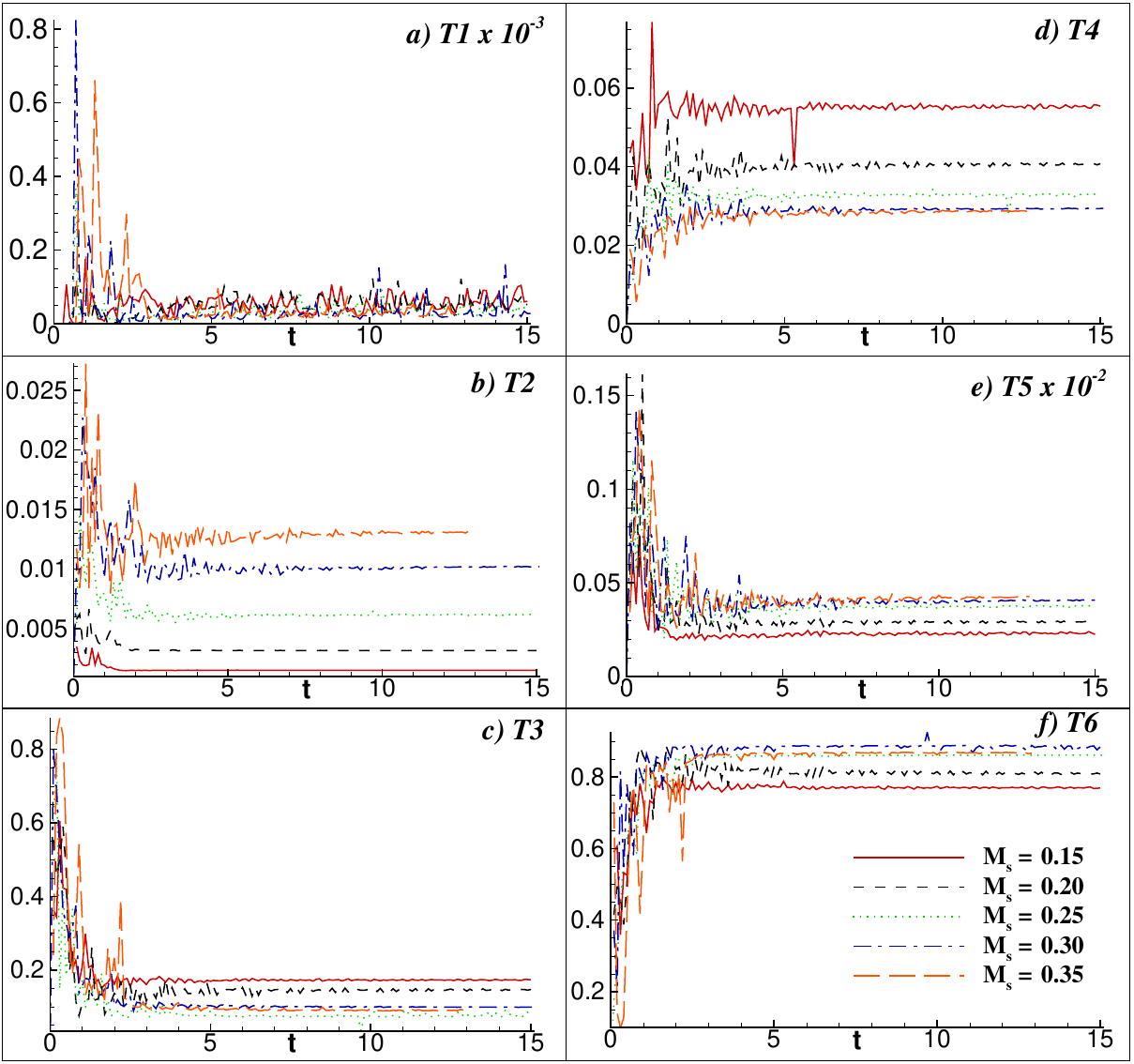}
\caption{Evolution of fractional contribution of maximum CETE budget terms, $(a) T1$, $(b) T2$, $(b) T3$, $(d) T4$, $(e) T5$, and $(f) T6$, for test cases with $M_s = 0.15$, 0.20, 0.25 and 0.30.}
\label{fig14}
\end{figure*}

Figure \ref{fig14} presents the temporal evolution of the dominant terms in the CETE for $M_s$ ranging from 0.15 to 0.35. Each term represents a distinct physical pathway for enstrophy generation, redistribution, or destruction in a compressible flow. The observed trends with increasing $M_s$ reflect how compressibility progressively alters the balance between shear-driven, baroclinic, and viscous–dilatational mechanisms.

The vortex stretching term, T1 shown in Fig. \ref{fig14}(a) increases with $M_s$, particularly at early times. This trend is expected because higher $M_s$ implies higher mean shear in the separated shear layer, three-dimensionality of vortical structures, and velocity gradients during transition and reattachment. At higher $M_s$, separation-induced transition occurs earlier and more abruptly, leading to rapid amplification and reorientation of vorticity vectors. Consequently, vortex stretching becomes increasingly important, especially during the initial growth phase of instabilities. Despite this increase, T1 remains relatively small compared to viscous-related terms, indicating that stretching alone does not dominate enstrophy evolution in this regime. The dilatation term, T2 increases monotonically with $M_s$ as shown in Fig. \ref{fig14}(b), reflecting the growing importance of compressibility effects. Larger values of $M_s$ implies higher velocity divergence, stronger local expansion and compression, enhanced interaction between vorticity and volumetric strain. In accelerating–decelerating flows typical of LPT suction surfaces, dilatation increasingly modulates enstrophy as $M_s$ rises. However, T2 remains secondary because dilatational effects primarily redistribute enstrophy rather than acting as a dominant source. The baroclinic term, T3, arising from misalignment of pressure and density gradients, shown in Fig. \ref{fig14}(c), decreases with $M_s$, except during very early stages. At lower $M_s$ separation bubbles are longer and more coherent, pressure and density gradients remain misaligned over extended regions, and baroclinic vorticity generation is sustained. As $M_s$ increases, transition occurs earlier, ensuing turbulent mixing homogenizes density and pressure fields more rapidly, and gradient misalignment weakens. This leads to a reduction in baroclinic enstrophy production at higher $M_s$. The early-time peak reflects the initial compressible acceleration around the leading edge before strong mixing sets in. The reduction in term T4 with $M_s$, as shown in Fig. \ref{fig14}(d), indicates that bulk-viscosity effects become less spatially misaligned with vorticity as turbulence intensifies. Increased mixing at higher $M_s$ smooths gradients responsible for this term. Thus, although compressibility increases, the coherence required to sustain T4 diminishes as the flow becomes more turbulent. The viscous diffusion term T5 shown in Fig. \ref{fig14}(e) increases with $M_s$ due to thicker boundary layers, enhanced velocity gradients, and stronger small-scale vorticity generation during early transition. Higher $M_s$ promotes more intense shear-layer breakdown, leading to greater enstrophy diffusion. Nevertheless, diffusion remains sub-dominant compared to viscous-stress–related production terms. The term T6 in Fig. \ref{fig14}(f) exhibits the largest contribution overall, increasing with $M_s$ and saturating between $M_s = 0.3$ and 0.35. This term encapsulates the direct coupling between density gradients and viscous stress divergence, making it uniquely important in compressible boundary layers. Its dominance indicates that enstrophy production in LPT flows is strongly governed by viscous–compressible interactions, rather than purely inviscid mechanisms. Higher $M_s$ intensify these interactions due to stronger density gradients and viscous stresses. The saturation at $M_s = 0.35$ suggests that beyond a certain $M_s$, enhanced mixing limits further growth of the misalignment between density and viscous stresses. The overall hierarchy of the CETE terms is: T6 >  T3 > T4 > T2 > T5 > T1 which highlights that: (i) viscous–compressible coupling is the primary driver of enstrophy evolution, (ii) baroclinic effects remain important, particularly at lower Mach numbers, and (iii) classical vortex stretching plays a comparatively minor role.

Increasing $M_s$ shifts enstrophy dynamics away from baroclinic and coherent-structure-driven mechanisms toward viscous–compressibility-dominated production pathways. This explains why separation bubbles shorten with $M_s$, momentum thickness and losses increase, and turbulence appears more distributed but less intermittently intense.

\section{Summary and conclusions \label{sec4}}

High-fidelity simulations are performed for inlet Mach numbers ranging from $M_s = 0.15$ to 0.35, to explore the influence of compressibility on separation-induced transition over a T106A low-pressure turbine blade (shown in Fig. \ref{fig1}) at elevated incidence. Numerical validation of the adopted dispersion relation preserving numerical methods is conducted in Fig. \ref{fig2} by comparisons of coefficient of pressure with the reported values in the DNS of Wissink \cite{wissink2006influence} and experiments of Stadmuller \cite{stadtmuller2001investigation}. The results demonstrate that increasing inlet Mach number fundamentally alters both the transition pathways and the mechanisms governing loss generation on the suction surface. Flow-field diagnostics in Figs. \ref{fig3} and \ref{fig4} highlight a clear shift in route of transition to turbulence with increasing Mach number. At low Mach numbers, transition is characterized by quasi-two-dimensional shear-layer roll-up, intermittent turbulent spots, and strong, localized turbulence production. As Mach number increases, transition becomes increasingly streak-dominated and bypass-like, with earlier onset, reduced coherence of spanwise rollers, and more spatially distributed turbulence. This change is reflected in space–time vorticity plot of Fig. \ref{fig5} and spectral analyses in Figs. \ref{fig6} and \ref{fig7}, which show a systematic redistribution of both temporal and spatial scales toward lower frequencies and larger structures. Compressibility enhances receptivity and promotes energy injection at large scales, rather than through gradual small-scale breakdown, as is evident from the spectral analysis depicted in Fig. \ref{fig8}. The turbulent cascade starts from low wavenumbers and extends downscale, which indicates a pathway close to bypass transition.  

Time-averaged coefficient of pressure in Fig. \ref{fig9} and skin-friction distribution in Fig. \ref{fig11} confirm the presence of leading-edge and trailing-edge separation bubbles across all Mach numbers. While higher Mach numbers lead to delayed separation and a reduced streamwise extent of these bubbles, this suppression of separation does not translate into lower losses. Instead, boundary-layer integral measures in Fig. \ref{fig12} reveal a substantial increase in suction-side momentum thickness with Mach number, indicating that compressibility enhances profile losses through intensified momentum deficit rather than prolonged separation.

Turbulent kinetic energy distributions in Fig. \ref{fig13} further indicate that lower Mach numbers produce higher peak turbulence intensities due to delayed but violent shear-layer breakdown, whereas higher Mach numbers yield lower peak TKE despite increased overall losses. The compressible enstrophy transport analysis shown in Fig. \ref{fig14} provides critical insight into this apparent contradiction. The results show that enstrophy evolution in compressible LPT flows is dominated by viscous–compressible coupling and baroclinic mechanisms, with vortex stretching playing a comparatively minor role. Increasing Mach number shifts the enstrophy balance toward viscous–compressibility-driven production, explaining the growth in momentum thickness and losses even as separation bubbles shorten.

Collectively, these findings demonstrate that compressibility modifies separation-induced transition in a nontrivial manner: it mitigates large-scale separation while simultaneously amplifying vorticity production and boundary-layer losses. The study underscores the limitations of separation length as a performance indicator and highlights the necessity of enstrophy-based diagnostics for developing physically informed transition models and loss predictions in low-pressure turbine design.

\begin{acknowledgments}
The authors would like to acknowledge the use of high-performance computing facility ARYABHATA at Indian Institute of Technology Dhanbad and computing cluster nautilus at Ecole Centrale de Nantes, France for all computations reported here. 
\end{acknowledgments}

\section*{Declaration of Interests}
The authors report no conflict of interest.

\section*{Data Availability}

The data will be made available upon reasonable request.



\bibliography{t106a}

\providecommand{\noopsort}[1]{}\providecommand{\singleletter}[1]{#1}%
\begin{thebibliography}{58}%
\makeatletter
\providecommand \@ifxundefined [1]{%
 \@ifx{#1\undefined}
}%
\providecommand \@ifnum [1]{%
 \ifnum #1\expandafter \@firstoftwo
 \else \expandafter \@secondoftwo
 \fi
}%
\providecommand \@ifx [1]{%
 \ifx #1\expandafter \@firstoftwo
 \else \expandafter \@secondoftwo
 \fi
}%
\providecommand \natexlab [1]{#1}%
\providecommand \enquote  [1]{``#1''}%
\providecommand \bibnamefont  [1]{#1}%
\providecommand \bibfnamefont [1]{#1}%
\providecommand \citenamefont [1]{#1}%
\providecommand \href@noop [0]{\@secondoftwo}%
\providecommand \href [0]{\begingroup \@sanitize@url \@href}%
\providecommand \@href[1]{\@@startlink{#1}\@@href}%
\providecommand \@@href[1]{\endgroup#1\@@endlink}%
\providecommand \@sanitize@url [0]{\catcode `\\12\catcode `\$12\catcode `\&12\catcode `\#12\catcode `\^12\catcode `\_12\catcode `\%12\relax}%
\providecommand \@@startlink[1]{}%
\providecommand \@@endlink[0]{}%
\providecommand \url  [0]{\begingroup\@sanitize@url \@url }%
\providecommand \@url [1]{\endgroup\@href {#1}{\urlprefix }}%
\providecommand \urlprefix  [0]{URL }%
\providecommand \Eprint [0]{\href }%
\providecommand \doibase [0]{http://dx.doi.org/}%
\providecommand \selectlanguage [0]{\@gobble}%
\providecommand \bibinfo  [0]{\@secondoftwo}%
\providecommand \bibfield  [0]{\@secondoftwo}%
\providecommand \translation [1]{[#1]}%
\providecommand \BibitemOpen [0]{}%
\providecommand \bibitemStop [0]{}%
\providecommand \bibitemNoStop [0]{.\EOS\space}%
\providecommand \EOS [0]{\spacefactor3000\relax}%
\providecommand \BibitemShut  [1]{\csname bibitem#1\endcsname}%
\let\auto@bib@innerbib\@empty
\bibitem [{\citenamefont {Halstead}\ \emph {et~al.}(1997)\citenamefont {Halstead}, \citenamefont {Wisler}, \citenamefont {Okiishi}, \citenamefont {Walker}, \citenamefont {Hodson},\ and\ \citenamefont {Shin}}]{halstead1997boundary}%
  \BibitemOpen
  \bibfield  {author} {\bibinfo {author} {\bibfnamefont {D.~E.}\ \bibnamefont {Halstead}}, \bibinfo {author} {\bibfnamefont {D.~C.}\ \bibnamefont {Wisler}}, \bibinfo {author} {\bibfnamefont {T.~H.}\ \bibnamefont {Okiishi}}, \bibinfo {author} {\bibfnamefont {G.~J.}\ \bibnamefont {Walker}}, \bibinfo {author} {\bibfnamefont {H.~P.}\ \bibnamefont {Hodson}}, \ and\ \bibinfo {author} {\bibfnamefont {H.~W.}\ \bibnamefont {Shin}},\ }\bibfield  {title} {\enquote {\bibinfo {title} {{Boundary layer development in axial compressors and turbines: Part 1 of 4—Composite picture}},}\ }\href@noop {} {\bibfield  {journal} {\bibinfo  {journal} {Journal of Turbomachinery}\ }\textbf {\bibinfo {volume} {119(1)}},\ \bibinfo {pages} {114--127} (\bibinfo {year} {1997})}\BibitemShut {NoStop}%
\bibitem [{\citenamefont {Hammer}, \citenamefont {Sandham},\ and\ \citenamefont {Sandberg}(2018)}]{hammer2018large}%
  \BibitemOpen
  \bibfield  {author} {\bibinfo {author} {\bibfnamefont {F.}~\bibnamefont {Hammer}}, \bibinfo {author} {\bibfnamefont {N.~D.}\ \bibnamefont {Sandham}}, \ and\ \bibinfo {author} {\bibfnamefont {R.~D.}\ \bibnamefont {Sandberg}},\ }\bibfield  {title} {\enquote {\bibinfo {title} {{Large Eddy Simulations of a Low-Pressure Turbine: Roughness Modeling and the Effects on Boundary Layer Transition and Losses}},}\ }in\ \href@noop {} {\emph {\bibinfo {booktitle} {Turbo Expo: Power for Land, Sea, and Air}}},\ Vol.\ \bibinfo {volume} {51005}\ (\bibinfo {organization} {American Society of Mechanical Engineers},\ \bibinfo {year} {2018})\ p.\ \bibinfo {pages} {V02BT41A014}\BibitemShut {NoStop}%
\bibitem [{\citenamefont {Banieghbal}\ \emph {et~al.}(1996)\citenamefont {Banieghbal}, \citenamefont {Curtis}, \citenamefont {Denton}, \citenamefont {Hodson}, \citenamefont {Hunstman}, \citenamefont {Schulte}, \citenamefont {Harvey},\ and\ \citenamefont {Steele}}]{banieghbal1996wake}%
  \BibitemOpen
  \bibfield  {author} {\bibinfo {author} {\bibfnamefont {M.~R.}\ \bibnamefont {Banieghbal}}, \bibinfo {author} {\bibfnamefont {E.~M.}\ \bibnamefont {Curtis}}, \bibinfo {author} {\bibfnamefont {J.~D.}\ \bibnamefont {Denton}}, \bibinfo {author} {\bibfnamefont {H.~P.}\ \bibnamefont {Hodson}}, \bibinfo {author} {\bibfnamefont {I.}~\bibnamefont {Hunstman}}, \bibinfo {author} {\bibfnamefont {V.}~\bibnamefont {Schulte}}, \bibinfo {author} {\bibfnamefont {N.~W.}\ \bibnamefont {Harvey}}, \ and\ \bibinfo {author} {\bibfnamefont {A.~B.}\ \bibnamefont {Steele}},\ }\bibfield  {title} {\enquote {\bibinfo {title} {{Wake passing in LP turbine blades}},}\ }in\ \href@noop {} {\emph {\bibinfo {booktitle} {AGARD CONFERENCE PROCEEDINGS AGARD CP}}}\ (\bibinfo {organization} {AGARD},\ \bibinfo {year} {1996})\ pp.\ \bibinfo {pages} {23--23}\BibitemShut {NoStop}%
\bibitem [{\citenamefont {Coull}\ and\ \citenamefont {Hodson}(2012)}]{coull2012predicting}%
  \BibitemOpen
  \bibfield  {author} {\bibinfo {author} {\bibfnamefont {J.~D.}\ \bibnamefont {Coull}}\ and\ \bibinfo {author} {\bibfnamefont {H.~P.}\ \bibnamefont {Hodson}},\ }\bibfield  {title} {\enquote {\bibinfo {title} {{Predicting the profile loss of high-lift low pressure turbines}},}\ }\href@noop {} {\bibfield  {journal} {\bibinfo  {journal} {Journal of Turbomachinery}\ }\textbf {\bibinfo {volume} {134(2)}} (\bibinfo {year} {2012})}\BibitemShut {NoStop}%
\bibitem [{\citenamefont {Mayle}(1991)}]{mayle1991role}%
  \BibitemOpen
  \bibfield  {author} {\bibinfo {author} {\bibfnamefont {R.~E.}\ \bibnamefont {Mayle}},\ }\bibfield  {title} {\enquote {\bibinfo {title} {The role of laminar-turbulent transition in gas turbine engines},}\ }in\ \href@noop {} {\emph {\bibinfo {booktitle} {Turbo Expo: Power for Land, Sea, and Air}}},\ Vol.\ \bibinfo {volume} {79023}\ (\bibinfo {organization} {American Society of Mechanical Engineers},\ \bibinfo {year} {1991})\ p.\ \bibinfo {pages} {V005T17A001}\BibitemShut {NoStop}%
\bibitem [{\citenamefont {Stadtm{\"u}ller}(2001)}]{stadtmuller2001investigation}%
  \BibitemOpen
  \bibfield  {author} {\bibinfo {author} {\bibfnamefont {P.}~\bibnamefont {Stadtm{\"u}ller}},\ }\bibfield  {title} {\enquote {\bibinfo {title} {{Investigation of wake-induced transition on the LP turbine cascade T106A-EIZ}},}\ }\href@noop {} {\bibfield  {journal} {\bibinfo  {journal} {DFG-Verbundprojekt Fo}\ }\textbf {\bibinfo {volume} {136}} (\bibinfo {year} {2001})}\BibitemShut {NoStop}%
\bibitem [{\citenamefont {Wissink}, \citenamefont {Rodi},\ and\ \citenamefont {Hodson}(2006)}]{wissink2006influence}%
  \BibitemOpen
  \bibfield  {author} {\bibinfo {author} {\bibfnamefont {J.~G.}\ \bibnamefont {Wissink}}, \bibinfo {author} {\bibfnamefont {W.}~\bibnamefont {Rodi}}, \ and\ \bibinfo {author} {\bibfnamefont {H.~P.}\ \bibnamefont {Hodson}},\ }\bibfield  {title} {\enquote {\bibinfo {title} {The influence of disturbances carried by periodically incoming wakes on the separating flow around a turbine blade},}\ }\href@noop {} {\bibfield  {journal} {\bibinfo  {journal} {International journal of heat and fluid flow}\ }\textbf {\bibinfo {volume} {27}},\ \bibinfo {pages} {721--729} (\bibinfo {year} {2006})}\BibitemShut {NoStop}%
\bibitem [{\citenamefont {Michelassi}\ \emph {et~al.}(2015)\citenamefont {Michelassi}, \citenamefont {Chen}, \citenamefont {Pichler},\ and\ \citenamefont {Sandberg}}]{michelassi2015compressible}%
  \BibitemOpen
  \bibfield  {author} {\bibinfo {author} {\bibfnamefont {V.}~\bibnamefont {Michelassi}}, \bibinfo {author} {\bibfnamefont {L.-W.}\ \bibnamefont {Chen}}, \bibinfo {author} {\bibfnamefont {R.}~\bibnamefont {Pichler}}, \ and\ \bibinfo {author} {\bibfnamefont {R.~D.}\ \bibnamefont {Sandberg}},\ }\bibfield  {title} {\enquote {\bibinfo {title} {{Compressible direct numerical simulation of low-pressure turbines—part II: effect of inflow disturbances}},}\ }\href@noop {} {\bibfield  {journal} {\bibinfo  {journal} {Journal of Turbomachinery}\ }\textbf {\bibinfo {volume} {137}},\ \bibinfo {pages} {071005} (\bibinfo {year} {2015})}\BibitemShut {NoStop}%
\bibitem [{\citenamefont {Volino}\ and\ \citenamefont {Hultgren}(2001)}]{volino2001measurements}%
  \BibitemOpen
  \bibfield  {author} {\bibinfo {author} {\bibfnamefont {R.~J.}\ \bibnamefont {Volino}}\ and\ \bibinfo {author} {\bibfnamefont {L.~S.}\ \bibnamefont {Hultgren}},\ }\bibfield  {title} {\enquote {\bibinfo {title} {Measurements in separated and transitional boundary layers under low-pressure turbine airfoil conditions},}\ }\href@noop {} {\bibfield  {journal} {\bibinfo  {journal} {J. Turbomach.}\ }\textbf {\bibinfo {volume} {123}},\ \bibinfo {pages} {189--197} (\bibinfo {year} {2001})}\BibitemShut {NoStop}%
\bibitem [{\citenamefont {Simoni}\ \emph {et~al.}(2015)\citenamefont {Simoni}, \citenamefont {Berrino}, \citenamefont {Ubaldi}, \citenamefont {Zunino},\ and\ \citenamefont {Bertini}}]{simoni2015off}%
  \BibitemOpen
  \bibfield  {author} {\bibinfo {author} {\bibfnamefont {D.}~\bibnamefont {Simoni}}, \bibinfo {author} {\bibfnamefont {M.}~\bibnamefont {Berrino}}, \bibinfo {author} {\bibfnamefont {M.}~\bibnamefont {Ubaldi}}, \bibinfo {author} {\bibfnamefont {P.}~\bibnamefont {Zunino}}, \ and\ \bibinfo {author} {\bibfnamefont {F.}~\bibnamefont {Bertini}},\ }\bibfield  {title} {\enquote {\bibinfo {title} {Off-design performance of a highly loaded low pressure turbine cascade under steady and unsteady incoming flow conditions},}\ }\href@noop {} {\bibfield  {journal} {\bibinfo  {journal} {Journal of Turbomachinery}\ }\textbf {\bibinfo {volume} {137}},\ \bibinfo {pages} {071009} (\bibinfo {year} {2015})}\BibitemShut {NoStop}%
\bibitem [{\citenamefont {Sengupta}\ and\ \citenamefont {Tucker}(2020{\natexlab{a}})}]{sengupta2020effects}%
  \BibitemOpen
  \bibfield  {author} {\bibinfo {author} {\bibfnamefont {A.}~\bibnamefont {Sengupta}}\ and\ \bibinfo {author} {\bibfnamefont {P.~G.}\ \bibnamefont {Tucker}},\ }\bibfield  {title} {\enquote {\bibinfo {title} {Effects of forced frequency oscillations and free stream turbulence on the separation-induced transition in pressure gradient dominated flows},}\ }\href@noop {} {\bibfield  {journal} {\bibinfo  {journal} {Physics of Fluids}\ }\textbf {\bibinfo {volume} {32}} (\bibinfo {year} {2020}{\natexlab{a}})}\BibitemShut {NoStop}%
\bibitem [{\citenamefont {Stadtm{\"u}ller}\ and\ \citenamefont {Fottner}(2001)}]{stadtmuller2001test}%
  \BibitemOpen
  \bibfield  {author} {\bibinfo {author} {\bibfnamefont {P.}~\bibnamefont {Stadtm{\"u}ller}}\ and\ \bibinfo {author} {\bibfnamefont {L.}~\bibnamefont {Fottner}},\ }\bibfield  {title} {\enquote {\bibinfo {title} {{A test case for the numerical investigation of wake passing effects on a highly loaded LP turbine cascade blade}},}\ }in\ \href@noop {} {\emph {\bibinfo {booktitle} {Turbo Expo: Power for Land, Sea, and Air}}},\ Vol.\ \bibinfo {volume} {78507}\ (\bibinfo {organization} {American Society of Mechanical Engineers},\ \bibinfo {year} {2001})\ p.\ \bibinfo {pages} {V001T03A015}\BibitemShut {NoStop}%
\bibitem [{\citenamefont {Wissink}(2003)}]{wissink2003dns}%
  \BibitemOpen
  \bibfield  {author} {\bibinfo {author} {\bibfnamefont {J.}~\bibnamefont {Wissink}},\ }\bibfield  {title} {\enquote {\bibinfo {title} {{DNS of separating, low Reynolds number flow in a turbine cascade with incoming wakes}},}\ }\href@noop {} {\bibfield  {journal} {\bibinfo  {journal} {International Journal of Heat and Fluid Flow}\ }\textbf {\bibinfo {volume} {24}},\ \bibinfo {pages} {626--635} (\bibinfo {year} {2003})}\BibitemShut {NoStop}%
\bibitem [{\citenamefont {Ranjan}, \citenamefont {Deshpande},\ and\ \citenamefont {Narasimha}(2014)}]{ranjan2014direct}%
  \BibitemOpen
  \bibfield  {author} {\bibinfo {author} {\bibfnamefont {R.}~\bibnamefont {Ranjan}}, \bibinfo {author} {\bibfnamefont {S.}~\bibnamefont {Deshpande}}, \ and\ \bibinfo {author} {\bibfnamefont {R.}~\bibnamefont {Narasimha}},\ }\bibfield  {title} {\enquote {\bibinfo {title} {Direct numerical simulation of compressible flow past a low pressure turbine blade at high incidence},}\ }in\ \href@noop {} {\emph {\bibinfo {booktitle} {Fluids Engineering Division Summer Meeting}}},\ Vol.\ \bibinfo {volume} {46216}\ (\bibinfo {organization} {American Society of Mechanical Engineers},\ \bibinfo {year} {2014})\ p.\ \bibinfo {pages} {V01AT02A010}\BibitemShut {NoStop}%
\bibitem [{\citenamefont {De~Vincentiis}\ \emph {et~al.}(2023)\citenamefont {De~Vincentiis}, \citenamefont {{\DH}urovi{\'c}}, \citenamefont {Lengani}, \citenamefont {Simoni}, \citenamefont {Pralits}, \citenamefont {Henningson},\ and\ \citenamefont {Hanifi}}]{de2023effects}%
  \BibitemOpen
  \bibfield  {author} {\bibinfo {author} {\bibfnamefont {L.}~\bibnamefont {De~Vincentiis}}, \bibinfo {author} {\bibfnamefont {K.}~\bibnamefont {{\DH}urovi{\'c}}}, \bibinfo {author} {\bibfnamefont {D.}~\bibnamefont {Lengani}}, \bibinfo {author} {\bibfnamefont {D.}~\bibnamefont {Simoni}}, \bibinfo {author} {\bibfnamefont {J.}~\bibnamefont {Pralits}}, \bibinfo {author} {\bibfnamefont {D.~S.}\ \bibnamefont {Henningson}}, \ and\ \bibinfo {author} {\bibfnamefont {A.}~\bibnamefont {Hanifi}},\ }\bibfield  {title} {\enquote {\bibinfo {title} {Effects of upstream wakes on the boundary layer over a low-pressure turbine blade},}\ }\href@noop {} {\bibfield  {journal} {\bibinfo  {journal} {Journal of turbomachinery}\ }\textbf {\bibinfo {volume} {145}},\ \bibinfo {pages} {051011} (\bibinfo {year} {2023})}\BibitemShut {NoStop}%
\bibitem [{\citenamefont {Pichler}\ \emph {et~al.}(2017)\citenamefont {Pichler}, \citenamefont {Michelassi}, \citenamefont {Sandberg},\ and\ \citenamefont {Ong}}]{pichler2017highly}%
  \BibitemOpen
  \bibfield  {author} {\bibinfo {author} {\bibfnamefont {R.}~\bibnamefont {Pichler}}, \bibinfo {author} {\bibfnamefont {V.}~\bibnamefont {Michelassi}}, \bibinfo {author} {\bibfnamefont {R.}~\bibnamefont {Sandberg}}, \ and\ \bibinfo {author} {\bibfnamefont {J.}~\bibnamefont {Ong}},\ }\bibfield  {title} {\enquote {\bibinfo {title} {{Highly resolved LES study of gap size effect on low-pressure turbine stage}},}\ }in\ \href@noop {} {\emph {\bibinfo {booktitle} {Turbo Expo: Power for Land, Sea, and Air}}},\ Vol.\ \bibinfo {volume} {50787}\ (\bibinfo {organization} {American Society of Mechanical Engineers},\ \bibinfo {year} {2017})\ p.\ \bibinfo {pages} {V02AT40A006}\BibitemShut {NoStop}%
\bibitem [{\citenamefont {Gross}, \citenamefont {Marks},\ and\ \citenamefont {Sondergaard}(2022)}]{gross2022numerical}%
  \BibitemOpen
  \bibfield  {author} {\bibinfo {author} {\bibfnamefont {A.}~\bibnamefont {Gross}}, \bibinfo {author} {\bibfnamefont {C.~R.}\ \bibnamefont {Marks}}, \ and\ \bibinfo {author} {\bibfnamefont {R.}~\bibnamefont {Sondergaard}},\ }\bibfield  {title} {\enquote {\bibinfo {title} {{Numerical Investigation of Freestream Turbulence Effect on Flow through Low-Pressure Turbine Cascade}},}\ }in\ \href@noop {} {\emph {\bibinfo {booktitle} {AIAA AVIATION 2022 Forum}}}\ (\bibinfo {year} {2022})\ p.\ \bibinfo {pages} {3412}\BibitemShut {NoStop}%
\bibitem [{\citenamefont {Fiore}\ \emph {et~al.}(2023)\citenamefont {Fiore}, \citenamefont {Gojon}, \citenamefont {S{\'a}ez-Mischlich},\ and\ \citenamefont {Gressier}}]{fiore2023t106}%
  \BibitemOpen
  \bibfield  {author} {\bibinfo {author} {\bibfnamefont {M.}~\bibnamefont {Fiore}}, \bibinfo {author} {\bibfnamefont {R.}~\bibnamefont {Gojon}}, \bibinfo {author} {\bibfnamefont {G.}~\bibnamefont {S{\'a}ez-Mischlich}}, \ and\ \bibinfo {author} {\bibfnamefont {J.}~\bibnamefont {Gressier}},\ }\bibfield  {title} {\enquote {\bibinfo {title} {{LES of the T106 low-pressure turbine: Spectral proper orthogonal decomposition of the flow based on a fluctuating energy norm}},}\ }\href@noop {} {\bibfield  {journal} {\bibinfo  {journal} {Computers \& Fluids}\ }\textbf {\bibinfo {volume} {252}},\ \bibinfo {pages} {105761} (\bibinfo {year} {2023})}\BibitemShut {NoStop}%
\bibitem [{\citenamefont {Sengupta}, \citenamefont {Vadlamani},\ and\ \citenamefont {Tucker}(2017)}]{sengupta2017roughness}%
  \BibitemOpen
  \bibfield  {author} {\bibinfo {author} {\bibfnamefont {A.}~\bibnamefont {Sengupta}}, \bibinfo {author} {\bibfnamefont {N.}~\bibnamefont {Vadlamani}}, \ and\ \bibinfo {author} {\bibfnamefont {P.~G.}\ \bibnamefont {Tucker}},\ }\bibfield  {title} {\enquote {\bibinfo {title} {Roughness induced transition in low pressure turbines},}\ }in\ \href@noop {} {\emph {\bibinfo {booktitle} {55th AIAA Aerospace Sciences Meeting}}}\ (\bibinfo {year} {2017})\ p.\ \bibinfo {pages} {0303}\BibitemShut {NoStop}%
\bibitem [{\citenamefont {Nakhchi}\ and\ \citenamefont {Rahmati}(2021)}]{nakhchi2021direct}%
  \BibitemOpen
  \bibfield  {author} {\bibinfo {author} {\bibfnamefont {M.}~\bibnamefont {Nakhchi}}\ and\ \bibinfo {author} {\bibfnamefont {M.}~\bibnamefont {Rahmati}},\ }\bibfield  {title} {\enquote {\bibinfo {title} {Direct numerical simulations of flutter instabilities over a vibrating turbine blade cascade},}\ }\href@noop {} {\bibfield  {journal} {\bibinfo  {journal} {Journal of Fluids and Structures}\ }\textbf {\bibinfo {volume} {104}},\ \bibinfo {pages} {103324} (\bibinfo {year} {2021})}\BibitemShut {NoStop}%
\bibitem [{\citenamefont {Sengupta}\ and\ \citenamefont {Tucker}(2020{\natexlab{b}})}]{sengupta2020effectsa}%
  \BibitemOpen
  \bibfield  {author} {\bibinfo {author} {\bibfnamefont {A.}~\bibnamefont {Sengupta}}\ and\ \bibinfo {author} {\bibfnamefont {P.}~\bibnamefont {Tucker}},\ }\bibfield  {title} {\enquote {\bibinfo {title} {Effects of forced frequency oscillations and unsteady wakes on the separation-induced transition in pressure gradient dominated flows},}\ }\href@noop {} {\bibfield  {journal} {\bibinfo  {journal} {Physics of Fluids}\ }\textbf {\bibinfo {volume} {32}} (\bibinfo {year} {2020}{\natexlab{b}})}\BibitemShut {NoStop}%
\bibitem [{\citenamefont {Michelassi}, \citenamefont {Wissink},\ and\ \citenamefont {Rodi}(2002)}]{michelassi2002analysis}%
  \BibitemOpen
  \bibfield  {author} {\bibinfo {author} {\bibfnamefont {V.}~\bibnamefont {Michelassi}}, \bibinfo {author} {\bibfnamefont {J.}~\bibnamefont {Wissink}}, \ and\ \bibinfo {author} {\bibfnamefont {W.}~\bibnamefont {Rodi}},\ }\bibfield  {title} {\enquote {\bibinfo {title} {Analysis of dns and les of flow in a low pressure turbine cascade with incoming wakes and comparison with experiments},}\ }\href@noop {} {\bibfield  {journal} {\bibinfo  {journal} {Flow, turbulence and combustion}\ }\textbf {\bibinfo {volume} {69}},\ \bibinfo {pages} {295--329} (\bibinfo {year} {2002})}\BibitemShut {NoStop}%
\bibitem [{\citenamefont {Fiore}\ and\ \citenamefont {Gourdain}(2021)}]{fiore2021reynolds}%
  \BibitemOpen
  \bibfield  {author} {\bibinfo {author} {\bibfnamefont {M.}~\bibnamefont {Fiore}}\ and\ \bibinfo {author} {\bibfnamefont {N.}~\bibnamefont {Gourdain}},\ }\bibfield  {title} {\enquote {\bibinfo {title} {Reynolds, mach, and freestream turbulence effects on the flow in a low-pressure turbine},}\ }\href@noop {} {\bibfield  {journal} {\bibinfo  {journal} {Journal of Turbomachinery}\ }\textbf {\bibinfo {volume} {143}},\ \bibinfo {pages} {101009} (\bibinfo {year} {2021})}\BibitemShut {NoStop}%
\bibitem [{\citenamefont {Sengupta}\ and\ \citenamefont {Sundaram}(2023)}]{sengupta2023compressibility}%
  \BibitemOpen
  \bibfield  {author} {\bibinfo {author} {\bibfnamefont {A.}~\bibnamefont {Sengupta}}\ and\ \bibinfo {author} {\bibfnamefont {P.}~\bibnamefont {Sundaram}},\ }\bibfield  {title} {\enquote {\bibinfo {title} {Compressibility effects on the flow past a t106a low-pressure turbine cascade},}\ }\href@noop {} {\bibfield  {journal} {\bibinfo  {journal} {Physics of Fluids}\ }\textbf {\bibinfo {volume} {35}} (\bibinfo {year} {2023})}\BibitemShut {NoStop}%
\bibitem [{\citenamefont {Duan}\ \emph {et~al.}(2023)\citenamefont {Duan}, \citenamefont {Qiao}, \citenamefont {Chen},\ and\ \citenamefont {Zhao}}]{duan2023effects}%
  \BibitemOpen
  \bibfield  {author} {\bibinfo {author} {\bibfnamefont {W.}~\bibnamefont {Duan}}, \bibinfo {author} {\bibfnamefont {W.}~\bibnamefont {Qiao}}, \bibinfo {author} {\bibfnamefont {W.}~\bibnamefont {Chen}}, \ and\ \bibinfo {author} {\bibfnamefont {X.}~\bibnamefont {Zhao}},\ }\bibfield  {title} {\enquote {\bibinfo {title} {Effects of freestream turbulence, reynolds number and mach number on the boundary layer in a low pressure turbine},}\ }\href@noop {} {\bibfield  {journal} {\bibinfo  {journal} {Journal of Thermal Science}\ }\textbf {\bibinfo {volume} {32}},\ \bibinfo {pages} {1393--1406} (\bibinfo {year} {2023})}\BibitemShut {NoStop}%
\bibitem [{\citenamefont {Pirozzoli}(2011)}]{pirozzoli2011numerical}%
  \BibitemOpen
  \bibfield  {author} {\bibinfo {author} {\bibfnamefont {S.}~\bibnamefont {Pirozzoli}},\ }\bibfield  {title} {\enquote {\bibinfo {title} {Numerical methods for high-speed flows},}\ }\href@noop {} {\bibfield  {journal} {\bibinfo  {journal} {Annual review of fluid mechanics}\ }\textbf {\bibinfo {volume} {43}},\ \bibinfo {pages} {163--194} (\bibinfo {year} {2011})}\BibitemShut {NoStop}%
\bibitem [{\citenamefont {Suman}\ \emph {et~al.}(2022)\citenamefont {Suman}, \citenamefont {Sundaram}, \citenamefont {Puttam}, \citenamefont {Sengupta},\ and\ \citenamefont {Sengupta}}]{suman2022novel}%
  \BibitemOpen
  \bibfield  {author} {\bibinfo {author} {\bibfnamefont {V.~K.}\ \bibnamefont {Suman}}, \bibinfo {author} {\bibfnamefont {P.}~\bibnamefont {Sundaram}}, \bibinfo {author} {\bibfnamefont {J.~K.}\ \bibnamefont {Puttam}}, \bibinfo {author} {\bibfnamefont {A.}~\bibnamefont {Sengupta}}, \ and\ \bibinfo {author} {\bibfnamefont {T.~K.}\ \bibnamefont {Sengupta}},\ }\bibfield  {title} {\enquote {\bibinfo {title} {{A novel compressible enstrophy transport equation-based analysis of instability during Magnus--Robins effects for high rotation rates}},}\ }\href@noop {} {\bibfield  {journal} {\bibinfo  {journal} {Physics of Fluids}\ }\textbf {\bibinfo {volume} {34}} (\bibinfo {year} {2022})}\BibitemShut {NoStop}%
\bibitem [{\citenamefont {Opoka}\ and\ \citenamefont {Hodson}(2007)}]{opoka2007transition}%
  \BibitemOpen
  \bibfield  {author} {\bibinfo {author} {\bibfnamefont {M.~M.}\ \bibnamefont {Opoka}}\ and\ \bibinfo {author} {\bibfnamefont {H.~P.}\ \bibnamefont {Hodson}},\ }\bibfield  {title} {\enquote {\bibinfo {title} {{Transition on the T106 LP Turbine Blade in the Presence of Moving Upstream Wakes and Downstream Potential Fields}},}\ }in\ \href@noop {} {\emph {\bibinfo {booktitle} {Turbo Expo: Power for Land, Sea, and Air}}},\ Vol.\ \bibinfo {volume} {47934}\ (\bibinfo {year} {2007})\ pp.\ \bibinfo {pages} {1091--1104}\BibitemShut {NoStop}%
\bibitem [{\citenamefont {Wissink}\ and\ \citenamefont {Rodi}(2004)}]{wissink2004dns}%
  \BibitemOpen
  \bibfield  {author} {\bibinfo {author} {\bibfnamefont {J.}~\bibnamefont {Wissink}}\ and\ \bibinfo {author} {\bibfnamefont {W.}~\bibnamefont {Rodi}},\ }\bibfield  {title} {\enquote {\bibinfo {title} {{DNS of a laminar separation bubble affected by free-stream disturbances}},}\ }in\ \href@noop {} {\emph {\bibinfo {booktitle} {Direct and Large-Eddy Simulation V: Proceedings of the fifth international ERCOFTAC Workshop on direct and large-eddy simulation held at the Munich University of Technology, August 27--29, 2003}}}\ (\bibinfo {organization} {Springer},\ \bibinfo {year} {2004})\ pp.\ \bibinfo {pages} {213--220}\BibitemShut {NoStop}%
\bibitem [{\citenamefont {Hirsch}(1990)}]{hirsch1990numerical}%
  \BibitemOpen
  \bibfield  {author} {\bibinfo {author} {\bibfnamefont {C.}~\bibnamefont {Hirsch}},\ }\bibfield  {title} {\enquote {\bibinfo {title} {{Numerical computation of internal and external flows}},}\ }\href@noop {} {\bibfield  {journal} {\bibinfo  {journal} {Computational methods for inviscid and viscous flows}\ }\textbf {\bibinfo {volume} {2}} (\bibinfo {year} {1990})}\BibitemShut {NoStop}%
\bibitem [{\citenamefont {Hoffmann}\ and\ \citenamefont {Chiang}(2000)}]{hoffmann2000computational}%
  \BibitemOpen
  \bibfield  {author} {\bibinfo {author} {\bibfnamefont {K.~A.}\ \bibnamefont {Hoffmann}}\ and\ \bibinfo {author} {\bibfnamefont {S.~T.}\ \bibnamefont {Chiang}},\ }\bibfield  {title} {\enquote {\bibinfo {title} {Computational fluid dynamics volume i},}\ }\href@noop {} {\bibfield  {journal} {\bibinfo  {journal} {Engineering education system}\ } (\bibinfo {year} {2000})}\BibitemShut {NoStop}%
\bibitem [{\citenamefont {Sagaut}\ \emph {et~al.}(2023)\citenamefont {Sagaut}, \citenamefont {Suman}, \citenamefont {Sundaram}, \citenamefont {Rajpoot}, \citenamefont {Bhumkar}, \citenamefont {Sengupta}, \citenamefont {Sengupta},\ and\ \citenamefont {Sengupta}}]{sagaut2023global}%
  \BibitemOpen
  \bibfield  {author} {\bibinfo {author} {\bibfnamefont {P.}~\bibnamefont {Sagaut}}, \bibinfo {author} {\bibfnamefont {V.~K.}\ \bibnamefont {Suman}}, \bibinfo {author} {\bibfnamefont {P.}~\bibnamefont {Sundaram}}, \bibinfo {author} {\bibfnamefont {M.~K.}\ \bibnamefont {Rajpoot}}, \bibinfo {author} {\bibfnamefont {Y.~G.}\ \bibnamefont {Bhumkar}}, \bibinfo {author} {\bibfnamefont {S.}~\bibnamefont {Sengupta}}, \bibinfo {author} {\bibfnamefont {A.}~\bibnamefont {Sengupta}}, \ and\ \bibinfo {author} {\bibfnamefont {T.~K.}\ \bibnamefont {Sengupta}},\ }\bibfield  {title} {\enquote {\bibinfo {title} {{Global spectral analysis: Review of numerical methods}},}\ }\href@noop {} {\bibfield  {journal} {\bibinfo  {journal} {Computers \& Fluids}\ ,\ \bibinfo {pages} {105915}} (\bibinfo {year} {2023})}\BibitemShut {NoStop}%
\bibitem [{\citenamefont {Sengupta}, \citenamefont {Sundaram},\ and\ \citenamefont {Sengupta}(2020)}]{sengupta2020nonmodal}%
  \BibitemOpen
  \bibfield  {author} {\bibinfo {author} {\bibfnamefont {A.}~\bibnamefont {Sengupta}}, \bibinfo {author} {\bibfnamefont {P.}~\bibnamefont {Sundaram}}, \ and\ \bibinfo {author} {\bibfnamefont {T.~K.}\ \bibnamefont {Sengupta}},\ }\bibfield  {title} {\enquote {\bibinfo {title} {{Nonmodal nonlinear route of transition to two-dimensional turbulence}},}\ }\href@noop {} {\bibfield  {journal} {\bibinfo  {journal} {Physical Review Research}\ }\textbf {\bibinfo {volume} {2}},\ \bibinfo {pages} {012033} (\bibinfo {year} {2020})}\BibitemShut {NoStop}%
\bibitem [{\citenamefont {Sundaram}\ \emph {et~al.}(2020)\citenamefont {Sundaram}, \citenamefont {Suman}, \citenamefont {Sengupta},\ and\ \citenamefont {Sengupta}}]{sundaram2020effects}%
  \BibitemOpen
  \bibfield  {author} {\bibinfo {author} {\bibfnamefont {P.}~\bibnamefont {Sundaram}}, \bibinfo {author} {\bibfnamefont {V.~K.}\ \bibnamefont {Suman}}, \bibinfo {author} {\bibfnamefont {A.}~\bibnamefont {Sengupta}}, \ and\ \bibinfo {author} {\bibfnamefont {T.~K.}\ \bibnamefont {Sengupta}},\ }\bibfield  {title} {\enquote {\bibinfo {title} {Effects of free stream excitation on the boundary layer over a semi-infinite flat plate},}\ }\href@noop {} {\bibfield  {journal} {\bibinfo  {journal} {Physics of Fluids}\ }\textbf {\bibinfo {volume} {32}} (\bibinfo {year} {2020})}\BibitemShut {NoStop}%
\bibitem [{\citenamefont {Sengupta}, \citenamefont {Gupta},\ and\ \citenamefont {Ubald}(2024)}]{sengupta2024separation}%
  \BibitemOpen
  \bibfield  {author} {\bibinfo {author} {\bibfnamefont {A.}~\bibnamefont {Sengupta}}, \bibinfo {author} {\bibfnamefont {N.}~\bibnamefont {Gupta}}, \ and\ \bibinfo {author} {\bibfnamefont {B.~N.}\ \bibnamefont {Ubald}},\ }\bibfield  {title} {\enquote {\bibinfo {title} {Separation-induced transition on a t106a blade under low and elevated free stream turbulence},}\ }\href@noop {} {\bibfield  {journal} {\bibinfo  {journal} {Physics of Fluids}\ }\textbf {\bibinfo {volume} {36}} (\bibinfo {year} {2024})}\BibitemShut {NoStop}%
\bibitem [{\citenamefont {Licari}\ and\ \citenamefont {Christensen}(2011)}]{licari2011modeling}%
  \BibitemOpen
  \bibfield  {author} {\bibinfo {author} {\bibfnamefont {A.}~\bibnamefont {Licari}}\ and\ \bibinfo {author} {\bibfnamefont {K.}~\bibnamefont {Christensen}},\ }\bibfield  {title} {\enquote {\bibinfo {title} {Modeling cumulative surface damage and assessing its impact on wall turbulence},}\ }\href@noop {} {\bibfield  {journal} {\bibinfo  {journal} {AIAA journal}\ }\textbf {\bibinfo {volume} {49}},\ \bibinfo {pages} {2305--2320} (\bibinfo {year} {2011})}\BibitemShut {NoStop}%
\bibitem [{\citenamefont {Adrian}(2007)}]{adrian2007hairpin}%
  \BibitemOpen
  \bibfield  {author} {\bibinfo {author} {\bibfnamefont {R.~J.}\ \bibnamefont {Adrian}},\ }\bibfield  {title} {\enquote {\bibinfo {title} {Hairpin vortex organization in wall turbulence},}\ }\href@noop {} {\bibfield  {journal} {\bibinfo  {journal} {Physics of fluids}\ }\textbf {\bibinfo {volume} {19}} (\bibinfo {year} {2007})}\BibitemShut {NoStop}%
\bibitem [{\citenamefont {Matsubara}\ and\ \citenamefont {Alfredsson}(2001)}]{matsubara2001disturbance}%
  \BibitemOpen
  \bibfield  {author} {\bibinfo {author} {\bibfnamefont {M.}~\bibnamefont {Matsubara}}\ and\ \bibinfo {author} {\bibfnamefont {P.~H.}\ \bibnamefont {Alfredsson}},\ }\bibfield  {title} {\enquote {\bibinfo {title} {{Disturbance growth in boundary layers subjected to free-stream turbulence}},}\ }\href@noop {} {\bibfield  {journal} {\bibinfo  {journal} {Journal of fluid mechanics}\ }\textbf {\bibinfo {volume} {430}},\ \bibinfo {pages} {149--168} (\bibinfo {year} {2001})}\BibitemShut {NoStop}%
\bibitem [{\citenamefont {Mohan}\ \emph {et~al.}(2021)\citenamefont {Mohan}, \citenamefont {Sameen}, \citenamefont {Srinivasan},\ and\ \citenamefont {Girimaji}}]{mohan2021influence}%
  \BibitemOpen
  \bibfield  {author} {\bibinfo {author} {\bibfnamefont {V.}~\bibnamefont {Mohan}}, \bibinfo {author} {\bibfnamefont {A.}~\bibnamefont {Sameen}}, \bibinfo {author} {\bibfnamefont {B.}~\bibnamefont {Srinivasan}}, \ and\ \bibinfo {author} {\bibfnamefont {S.~S.}\ \bibnamefont {Girimaji}},\ }\bibfield  {title} {\enquote {\bibinfo {title} {Influence of knudsen and mach numbers on kelvin-helmholtz instability},}\ }\href@noop {} {\bibfield  {journal} {\bibinfo  {journal} {Physical Review E}\ }\textbf {\bibinfo {volume} {103}},\ \bibinfo {pages} {053104} (\bibinfo {year} {2021})}\BibitemShut {NoStop}%
\bibitem [{\citenamefont {Doering}\ and\ \citenamefont {Gibbon}(1995)}]{doering1995applied}%
  \BibitemOpen
  \bibfield  {author} {\bibinfo {author} {\bibfnamefont {C.~R.}\ \bibnamefont {Doering}}\ and\ \bibinfo {author} {\bibfnamefont {J.~D.}\ \bibnamefont {Gibbon}},\ }\href@noop {} {\emph {\bibinfo {title} {{Applied analysis of the Navier-Stokes equations}}}},\ \bibinfo {number} {12}\ (\bibinfo  {publisher} {Cambridge university press},\ \bibinfo {year} {1995})\BibitemShut {NoStop}%
\bibitem [{\citenamefont {Klebanoff}, \citenamefont {Tidstrom},\ and\ \citenamefont {Sargent}(1962)}]{klebanoff1962three}%
  \BibitemOpen
  \bibfield  {author} {\bibinfo {author} {\bibfnamefont {P.~S.}\ \bibnamefont {Klebanoff}}, \bibinfo {author} {\bibfnamefont {K.}~\bibnamefont {Tidstrom}}, \ and\ \bibinfo {author} {\bibfnamefont {L.}~\bibnamefont {Sargent}},\ }\bibfield  {title} {\enquote {\bibinfo {title} {The three-dimensional nature of boundary-layer instability},}\ }\href@noop {} {\bibfield  {journal} {\bibinfo  {journal} {Journal of Fluid Mechanics}\ }\textbf {\bibinfo {volume} {12}},\ \bibinfo {pages} {1--34} (\bibinfo {year} {1962})}\BibitemShut {NoStop}%
\bibitem [{\citenamefont {Sandham}(2016)}]{sandham2016effects}%
  \BibitemOpen
  \bibfield  {author} {\bibinfo {author} {\bibfnamefont {N.~D.}\ \bibnamefont {Sandham}},\ }\bibfield  {title} {\enquote {\bibinfo {title} {Effects of compressibility and shock-wave interactions on turbulent shear flows},}\ }\href@noop {} {\bibfield  {journal} {\bibinfo  {journal} {Flow, Turbulence and Combustion}\ }\textbf {\bibinfo {volume} {97}},\ \bibinfo {pages} {1--25} (\bibinfo {year} {2016})}\BibitemShut {NoStop}%
\bibitem [{\citenamefont {Sieverding}(1985)}]{sieverding1985recent}%
  \BibitemOpen
  \bibfield  {author} {\bibinfo {author} {\bibfnamefont {C.}~\bibnamefont {Sieverding}},\ }\bibfield  {title} {\enquote {\bibinfo {title} {Recent progress in the understanding of basic aspects of secondary flows in turbine blade passages},}\ }\href@noop {} {\bibfield  {journal} {\bibinfo  {journal} {Journal of Engineering for Gas Turbines and Power}\ }\textbf {\bibinfo {volume} {107}},\ \bibinfo {pages} {248--257} (\bibinfo {year} {1985})}\BibitemShut {NoStop}%
\bibitem [{\citenamefont {Tennekes}\ and\ \citenamefont {Lumley}(1972)}]{tennekes1972first}%
  \BibitemOpen
  \bibfield  {author} {\bibinfo {author} {\bibfnamefont {H.}~\bibnamefont {Tennekes}}\ and\ \bibinfo {author} {\bibfnamefont {J.~L.}\ \bibnamefont {Lumley}},\ }\href@noop {} {\emph {\bibinfo {title} {A first course in turbulence}}}\ (\bibinfo  {publisher} {MIT press},\ \bibinfo {year} {1972})\BibitemShut {NoStop}%
\bibitem [{\citenamefont {Lele}(1994)}]{lele1994compressibility}%
  \BibitemOpen
  \bibfield  {author} {\bibinfo {author} {\bibfnamefont {S.~K.}\ \bibnamefont {Lele}},\ }\bibfield  {title} {\enquote {\bibinfo {title} {Compressibility effects on turbulence},}\ }\href@noop {} {\bibfield  {journal} {\bibinfo  {journal} {Annual review of fluid mechanics}\ }\textbf {\bibinfo {volume} {26}},\ \bibinfo {pages} {211--254} (\bibinfo {year} {1994})}\BibitemShut {NoStop}%
\bibitem [{\citenamefont {Garai}\ \emph {et~al.}(2015)\citenamefont {Garai}, \citenamefont {Diosady}, \citenamefont {Murman},\ and\ \citenamefont {Madavan}}]{garai2015dns}%
  \BibitemOpen
  \bibfield  {author} {\bibinfo {author} {\bibfnamefont {A.}~\bibnamefont {Garai}}, \bibinfo {author} {\bibfnamefont {L.}~\bibnamefont {Diosady}}, \bibinfo {author} {\bibfnamefont {S.}~\bibnamefont {Murman}}, \ and\ \bibinfo {author} {\bibfnamefont {N.}~\bibnamefont {Madavan}},\ }\bibfield  {title} {\enquote {\bibinfo {title} {Dns of flow in a low-pressure turbine cascade using a discontinuous-galerkin spectral-element method},}\ }in\ \href@noop {} {\emph {\bibinfo {booktitle} {Turbo Expo: Power for Land, Sea, and Air}}},\ Vol.\ \bibinfo {volume} {56642}\ (\bibinfo {organization} {American Society of Mechanical Engineers},\ \bibinfo {year} {2015})\ p.\ \bibinfo {pages} {V02BT39A023}\BibitemShut {NoStop}%
\bibitem [{\citenamefont {Denton}(1993)}]{denton1993loss}%
  \BibitemOpen
  \bibfield  {author} {\bibinfo {author} {\bibfnamefont {J.~D.}\ \bibnamefont {Denton}},\ }\href@noop {} {\emph {\bibinfo {title} {Loss mechanisms in turbomachines}}},\ Vol.\ \bibinfo {volume} {78897}\ (\bibinfo  {publisher} {American Society of Mechanical Engineers},\ \bibinfo {year} {1993})\BibitemShut {NoStop}%
\bibitem [{\citenamefont {Schlichting}\ and\ \citenamefont {Kestin}(1961)}]{schlichting1961boundary}%
  \BibitemOpen
  \bibfield  {author} {\bibinfo {author} {\bibfnamefont {H.}~\bibnamefont {Schlichting}}\ and\ \bibinfo {author} {\bibfnamefont {J.}~\bibnamefont {Kestin}},\ }\href@noop {} {\emph {\bibinfo {title} {Boundary layer theory}}},\ Vol.\ \bibinfo {volume} {121}\ (\bibinfo  {publisher} {Springer},\ \bibinfo {year} {1961})\BibitemShut {NoStop}%
\bibitem [{\citenamefont {Curtis}\ \emph {et~al.}(1997)\citenamefont {Curtis}, \citenamefont {Hodson}, \citenamefont {Banieghbal}, \citenamefont {Denton}, \citenamefont {Howell},\ and\ \citenamefont {Harvey}}]{curtis1997development}%
  \BibitemOpen
  \bibfield  {author} {\bibinfo {author} {\bibfnamefont {E.}~\bibnamefont {Curtis}}, \bibinfo {author} {\bibfnamefont {H.}~\bibnamefont {Hodson}}, \bibinfo {author} {\bibfnamefont {M.}~\bibnamefont {Banieghbal}}, \bibinfo {author} {\bibfnamefont {J.}~\bibnamefont {Denton}}, \bibinfo {author} {\bibfnamefont {R.}~\bibnamefont {Howell}}, \ and\ \bibinfo {author} {\bibfnamefont {N.}~\bibnamefont {Harvey}},\ }\bibfield  {title} {\enquote {\bibinfo {title} {{Development of blade profiles for low-pressure turbine applications}},}\ }\href@noop {} {\bibfield  {journal} {\bibinfo  {journal} {Journal of Turbomachinery}\ }\textbf {\bibinfo {volume} {119(3)}},\ \bibinfo {pages} {531--538} (\bibinfo {year} {1997})}\BibitemShut {NoStop}%
\bibitem [{\citenamefont {Alam}\ and\ \citenamefont {Sandham}(2000)}]{alam2000direct}%
  \BibitemOpen
  \bibfield  {author} {\bibinfo {author} {\bibfnamefont {M.}~\bibnamefont {Alam}}\ and\ \bibinfo {author} {\bibfnamefont {N.~D.}\ \bibnamefont {Sandham}},\ }\bibfield  {title} {\enquote {\bibinfo {title} {{Direct numerical simulation of ‘short’laminar separation bubbles with turbulent reattachment}},}\ }\href@noop {} {\bibfield  {journal} {\bibinfo  {journal} {Journal of Fluid Mechanics}\ }\textbf {\bibinfo {volume} {410}},\ \bibinfo {pages} {1--28} (\bibinfo {year} {2000})}\BibitemShut {NoStop}%
\bibitem [{\citenamefont {Sengupta}\ \emph {et~al.}(2018)\citenamefont {Sengupta}, \citenamefont {Suman}, \citenamefont {Sengupta},\ and\ \citenamefont {Bhaumik}}]{sengupta2018enstrophy}%
  \BibitemOpen
  \bibfield  {author} {\bibinfo {author} {\bibfnamefont {A.}~\bibnamefont {Sengupta}}, \bibinfo {author} {\bibfnamefont {V.~K.}\ \bibnamefont {Suman}}, \bibinfo {author} {\bibfnamefont {T.~K.}\ \bibnamefont {Sengupta}}, \ and\ \bibinfo {author} {\bibfnamefont {S.}~\bibnamefont {Bhaumik}},\ }\bibfield  {title} {\enquote {\bibinfo {title} {An enstrophy-based linear and nonlinear receptivity theory},}\ }\href@noop {} {\bibfield  {journal} {\bibinfo  {journal} {Physics of Fluids}\ }\textbf {\bibinfo {volume} {30}} (\bibinfo {year} {2018})}\BibitemShut {NoStop}%
\bibitem [{\citenamefont {Sengupta}, \citenamefont {Kumar},\ and\ \citenamefont {Kumar}(2025)}]{sengupta2025bifurcation}%
  \BibitemOpen
  \bibfield  {author} {\bibinfo {author} {\bibfnamefont {A.}~\bibnamefont {Sengupta}}, \bibinfo {author} {\bibfnamefont {S.}~\bibnamefont {Kumar}}, \ and\ \bibinfo {author} {\bibfnamefont {S.}~\bibnamefont {Kumar}},\ }\bibfield  {title} {\enquote {\bibinfo {title} {Bifurcation analysis of compressible flow past a rotating cylinder for high rotation rate},}\ }\href@noop {} {\bibfield  {journal} {\bibinfo  {journal} {Physics of Fluids}\ }\textbf {\bibinfo {volume} {37}} (\bibinfo {year} {2025})}\BibitemShut {NoStop}%
\bibitem [{\citenamefont {Joshi}\ \emph {et~al.}(2025)\citenamefont {Joshi}, \citenamefont {Sengupta}, \citenamefont {Ajanif},\ and\ \citenamefont {Lestandi}}]{joshi2025comparing}%
  \BibitemOpen
  \bibfield  {author} {\bibinfo {author} {\bibfnamefont {B.}~\bibnamefont {Joshi}}, \bibinfo {author} {\bibfnamefont {A.}~\bibnamefont {Sengupta}}, \bibinfo {author} {\bibfnamefont {Y.}~\bibnamefont {Ajanif}}, \ and\ \bibinfo {author} {\bibfnamefont {L.}~\bibnamefont {Lestandi}},\ }\bibfield  {title} {\enquote {\bibinfo {title} {Comparing the highly-resolved onset of rayleigh--taylor and kelvin--helmholtz rayleigh--taylor instabilities},}\ }\href@noop {} {\bibfield  {journal} {\bibinfo  {journal} {European Journal of Mechanics-B/Fluids}\ ,\ \bibinfo {pages} {204382}} (\bibinfo {year} {2025})}\BibitemShut {NoStop}%
\bibitem [{\citenamefont {Sengupta}(2025)}]{sengupta2025compressible}%
  \BibitemOpen
  \bibfield  {author} {\bibinfo {author} {\bibfnamefont {A.}~\bibnamefont {Sengupta}},\ }\bibfield  {title} {\enquote {\bibinfo {title} {Compressible enstrophy transport for flow in a low-pressure turbine with unsteady wakes impinging at the inflow},}\ }in\ \href@noop {} {\emph {\bibinfo {booktitle} {Computational Fluid Dynamics: Novel Numerical and Computational Approaches: Methodology and Numerics}}}\ (\bibinfo  {publisher} {Springer},\ \bibinfo {year} {2025})\ pp.\ \bibinfo {pages} {59--85}\BibitemShut {NoStop}%
\bibitem [{\citenamefont {Sengupta}\ and\ \citenamefont {Shandilya}(2024)}]{sengupta2024thermal}%
  \BibitemOpen
  \bibfield  {author} {\bibinfo {author} {\bibfnamefont {A.}~\bibnamefont {Sengupta}}\ and\ \bibinfo {author} {\bibfnamefont {N.}~\bibnamefont {Shandilya}},\ }\bibfield  {title} {\enquote {\bibinfo {title} {Thermal optimization of shock-induced separation in a natural laminar airfoil operating at off-design conditions},}\ }\href@noop {} {\bibfield  {journal} {\bibinfo  {journal} {Physics of Fluids}\ }\textbf {\bibinfo {volume} {36}} (\bibinfo {year} {2024})}\BibitemShut {NoStop}%
\bibitem [{\citenamefont {Lin}\ and\ \citenamefont {Wu}(2022)}]{lin2022physical}%
  \BibitemOpen
  \bibfield  {author} {\bibinfo {author} {\bibfnamefont {L.}~\bibnamefont {Lin}}\ and\ \bibinfo {author} {\bibfnamefont {Y.}~\bibnamefont {Wu}},\ }\bibfield  {title} {\enquote {\bibinfo {title} {Physical origin of vortex stretching and twisting: Viscous or inertial forces},}\ }\href@noop {} {\bibfield  {journal} {\bibinfo  {journal} {Physics of Fluids}\ }\textbf {\bibinfo {volume} {34}} (\bibinfo {year} {2022})}\BibitemShut {NoStop}%
\bibitem [{\citenamefont {Sarkar}\ \emph {et~al.}(1991)\citenamefont {Sarkar}, \citenamefont {Erlebacher}, \citenamefont {Hussaini},\ and\ \citenamefont {Kreiss}}]{sarkar1991analysis}%
  \BibitemOpen
  \bibfield  {author} {\bibinfo {author} {\bibfnamefont {S.}~\bibnamefont {Sarkar}}, \bibinfo {author} {\bibfnamefont {G.}~\bibnamefont {Erlebacher}}, \bibinfo {author} {\bibfnamefont {M.~Y.}\ \bibnamefont {Hussaini}}, \ and\ \bibinfo {author} {\bibfnamefont {H.~O.}\ \bibnamefont {Kreiss}},\ }\bibfield  {title} {\enquote {\bibinfo {title} {The analysis and modelling of dilatational terms in compressible turbulence},}\ }\href@noop {} {\bibfield  {journal} {\bibinfo  {journal} {Journal of Fluid Mechanics}\ }\textbf {\bibinfo {volume} {227}},\ \bibinfo {pages} {473--493} (\bibinfo {year} {1991})}\BibitemShut {NoStop}%
\bibitem [{\citenamefont {Jahanbakhshi}, \citenamefont {Vaghefi},\ and\ \citenamefont {Madnia}(2015)}]{jahanbakhshi2015baroclinic}%
  \BibitemOpen
  \bibfield  {author} {\bibinfo {author} {\bibfnamefont {R.}~\bibnamefont {Jahanbakhshi}}, \bibinfo {author} {\bibfnamefont {N.~S.}\ \bibnamefont {Vaghefi}}, \ and\ \bibinfo {author} {\bibfnamefont {C.~K.}\ \bibnamefont {Madnia}},\ }\bibfield  {title} {\enquote {\bibinfo {title} {Baroclinic vorticity generation near the turbulent/non-turbulent interface in a compressible shear layer},}\ }\href@noop {} {\bibfield  {journal} {\bibinfo  {journal} {Physics of Fluids}\ }\textbf {\bibinfo {volume} {27}} (\bibinfo {year} {2015})}\BibitemShut {NoStop}%
\end{thebibliography}%


\end{document}